\pdfoutput=1

\documentclass[aps,prl,reprint,english,superscriptaddress,amsmath,amssymb,floatfix,longbibliography]{revtex4-2}
\usepackage{blindtext}
\usepackage{graphicx}
\usepackage{amssymb}
\usepackage{amssymb,amsmath}
\usepackage{blindtext}
\usepackage{url}
\usepackage{ mathrsfs }
\usepackage{mathtools}
\usepackage{gensymb}
\usepackage{ textcomp }

\usepackage[normalem]{ulem}
\usepackage{color}

\begin{document}
\title{Multiscale Microtubule Dynamics in Active Nematics}
\author{Linnea M. Lemma\textsuperscript{1,2}, Michael M. Norton\textsuperscript{1}, Alexandra M. Tayar\textsuperscript{2}, Stephen J. DeCamp\textsuperscript{1}, S. Ali Aghvami\textsuperscript{1}, Seth Fraden\textsuperscript{1}, Michael F. Hagan\textsuperscript{1} and Zvonimir Dogic}
\email{zdogic@physics.ucsb.edu}
\affiliation{Department of Physics, Brandeis University, Waltham, MA USA 02474}, \affiliation{Department of Physics, University of California at Santa Barbara, Santa Barbara, CA USA 93106}
\date{August 16, 2021}

\begin{abstract}
In microtubule-based active nematics, motor-driven extensile motion of microtubule bundles powers chaotic large-scale dynamics. We quantify the interfilament sliding motion both in isolated bundles and in a dense active nematic. The extension speed of an isolated microtubule pair is comparable to the  molecular motor stepping speed. In contrast, the net extension in dense 2D active nematics is significantly slower; the interfilament sliding speeds are widely distributed about the average and the filaments exhibit both contractile and extensile relative motion. These measurements highlight the challenge of connecting the extension rate of isolated bundles to the multi-motor and multi-filament interactions present in a dense 2D active nematic. They also provide quantitative data that is essential for building multiscale models.

\end{abstract}

\maketitle

Driven by the continuous injection of energy through the motion of microscopic constituents, active nematics exhibit chaotic flows that are tightly coupled to the continuous creation and annihilation of motile topological defects~\cite{MarchettiReview, Narayan, LLCIgor, EpitheliaNematic, SanoNematic}. Such dynamics are described by continuum models in which the activity is introduced through a phenomenological active stress~\cite{Shi, ThampiTurbulence, DefectsGiomi,  SurajDefectUnbinding, Putzig2016SoftMatter, Oza_2016}. Experimental work  characterized the large-scale non-equilibrium dynamics of active nematics~\cite{DeCamp2015, Pau2017, GardelActinNematic, VortexPaperLemma, AmandaTan}. In comparison, little is known about how the microscopic constituents determine the coarse-grained dynamics, especially for cytoskeletal active nematics~\cite{ShelleyReview, BezSebastian}. For instance, while active stress is the defining feature of active nematics, predicting its  magnitude or sign in terms of microscopic dynamics remains a significant challenge~\cite{KruseContracting2000, Nedelec2001, GardelLenzContractile, NeedlemanDogicReview, Peter-eLife2015}.  

 Single-molecule experiments elucidated the dependence of the kinesin stepping speed on both the magnitude and the direction of the load force it experiences~\cite{HowardVale, StochasticKinesin, SchnitzerForce, KinesinMultiMotor, HowardPRL}. Consequently, determining the motor load and speed in active nematics is important for informing the microscopic models. However, measuring the relative speed of a motor with the respect to the filament it is stepping along is not possible in dense active nematics. In the absence of such data, the relative motion of filaments provide valuable insight into microscopic dynamics. For example, several models predict that filaments will slide past each other at a constant speed determined by the molecular motors~\cite{KrusePolar, GaoBetterton, BettertonSoftMatter, BezSebastian, FreyModel}. To test these predictions, we characterize the multiscale dynamics of microtubule-based active nematics. We find that dilute microtubule pairs extend at a constant speed, similar to the stepping dynamics of an isolated motor. In contrast to both the dilute bundles and theoretical predictions, the motion of microtubules in a dense nematic is more complex. The filament extension rate is both significantly slower and more widely distributed about the average when compared to the dynamics of isolated bundles. These effects must be accounted for to develop accurate multiscale models that predict the chaotic dynamics of active nematics from the properties of the microscopic constituents. 
 
 \begin{figure}

\includegraphics[width=8.6cm]{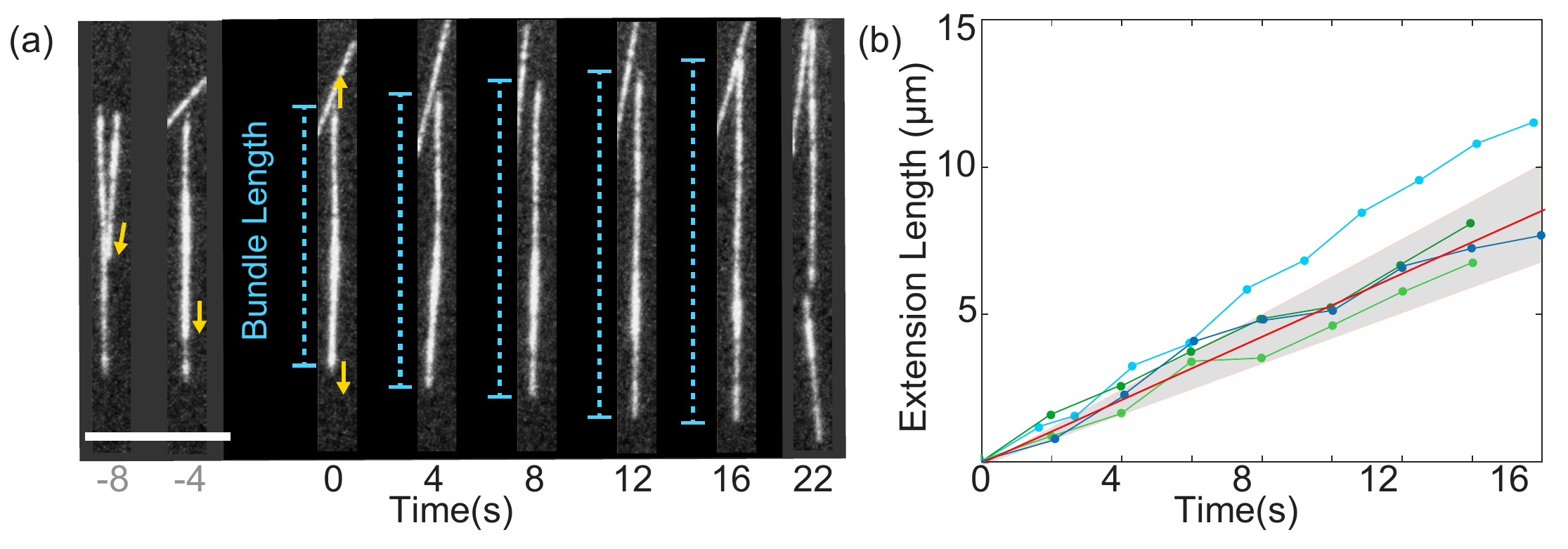}
\caption{\small \textbf{Dynamics of extensile bundles.} \textbf{(a)} Time series of two oppositely polarized microtubules cross-linked by kinesin motor clusters. Depletion induces bundle formation at t=0 s. Thereafter, kinesin clusters generate bundle extension. Scale bar, 10 $\mu$m. \textbf{(b)} Bundle length over time for four representative events. The red line indicates the average velocity of eight events, with shaded standard deviation. Average extension velocity $v=540 \pm 100$ nm/s. }
\label{Sliding}

\end{figure}

Extensile microtubule bundles are an essential building block of diverse active matter systems~\cite{FeodorFriction, Sanchez2012, Henkin2014, Pooja-arxiv, DuclosAdkins3D}, yet their dynamics have not been quantified. We adsorbed dilute microtubules onto a surfactant stabilized oil-water interface in the presence of kinesin motor clusters, a depletion agent, and an ATP regeneration system [Fig. 1(a)]. When two microtubules bundled together, the kinesin clusters linked adjacent filaments and moved toward their plus ends. For pairs with opposite polarity, motors stepping generated bundle extension [Fig. 1(b), Supp. Fig. 1]. Notably, the bundle length  increased linearly with time, allowing us to extract the extension speed from the slope. The average extension speed was $\langle v_e \rangle = 540 \pm 100$ nm/s. For comparison, a single kinesin-1 motor steps with an average speed of $\langle v \rangle=610\pm160$ nm/s~\cite{Berliner}, which extrapolates to a bundle extension speed of 1.2 $\mu$m/s. This discrepancy can have different causes. For example, the crowding agent or cluster structure could change the angle and the force load with which motors attach to microtubules~\cite{HowardPRL}. Diffusional dynamics suggest that the frictional force of interfilament sliding is $10^{-1}$ pN [Supp. Info.]. This is large compared to the $10^{-5}$ pN drag force on a 100 nm bead, but small when compared to the 5 pN force at which kinesin motors stall~\cite{FeodorFriction, Ward2015,SchnitzerForce}.

\begin{figure}

\includegraphics[width=8.6cm]{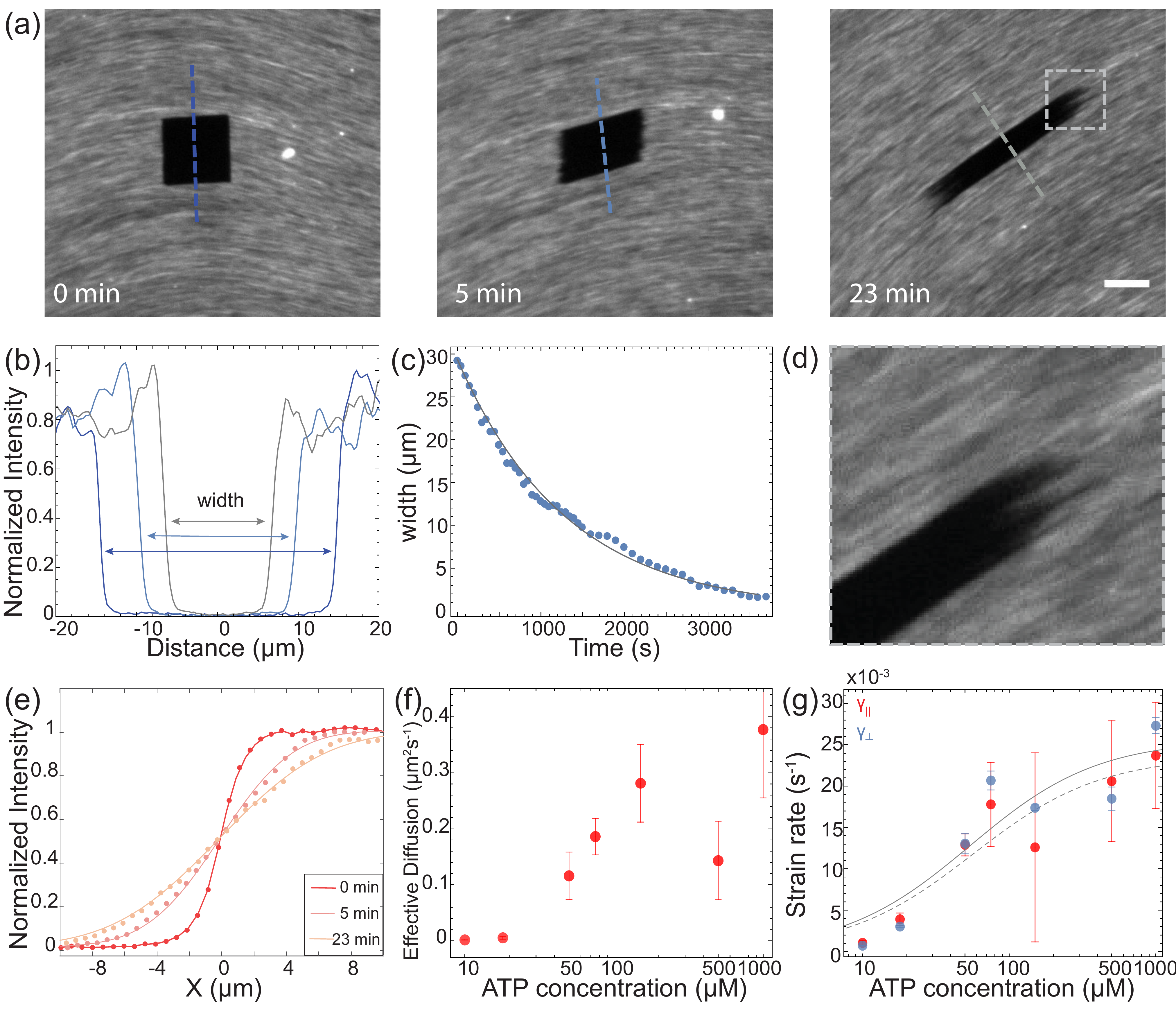}
\caption{\small \textbf{Analysis of photobleached active nematics.} \textbf{(a)} Time evolution of a photobleached region within a nematic assembled at 10 $\mu$M ATP. Scale bar, 20 $\mu$m. \textbf{(b)} Intensity profiles $I_{\perp}$ along the dotted lines in (a) and averaged across the length of the photobleached region. \textbf{(c)} The width of the photobleached region plotted over time, fitted to $w=ae^{-t/\tau}$.  \textbf{(d)} Initially sharp, the edge of the bleached square along the director roughens over time. \textbf{(e)} Intensity profiles $I_{||}(x)$ along the director averaged across the edge of the bleached region (points) and the prediction from the model of diffusion in Eq. \ref{diffusion} (lines). \textbf{(f)} The effective diffusion coefficient $D_{\textrm{bleach}}$ extracted for Eq. \ref{diffusion} versus ATP concentration. \textbf{(g)} The strain rate as a function of ATP concentration.  Strain rates extracted from the exponential fit $\gamma_{\perp}=1/\tau$ (blue) and from a fit to the diffusion-convection model Eq.\ref{diffusion} (red). The grey lines are fits to the Michaelis-Menten equation $\gamma=\gamma_{\textrm{max}} [\textrm{ATP}]/(K_{\textrm{M}}-[\textrm{ATP}])$ with $\gamma_{\perp,\textrm{max}}=0.026$ $\textrm{s}^{-1}$ and $K_{\perp,\textrm{M}}=54$ $\mu$M (blue, dashed) and $\gamma_{||,\textrm{max}}=0.027$ $\textrm{s}^{-1}$ and $K_{||,\textrm{M}}=46$ $\mu$M (red, solid). Error bars in (f) and (g) are the standard errors of multiple measurements.}

\label{FRAP}

\end{figure}

The extensional bundle motion drives the large-scale dynamics of active nematics. To relate the two phenomena, we next measured the dynamics of microtubules in dense active nematics. Sedimenting a high-density of extensile bundles onto a surfactant stabilized oil-water interface yielded an active nematic with local orientational order that was imaged using fluorescence microscopy. Nematic regions were labeled by photobleaching a {\raise.17ex\hbox{$\scriptstyle\mathtt{\sim}$}}50 $\mu\textrm{m}^2$ square in a defect-free, uniformly aligned region. Controls confirmed that the laser did not ablate the bleached microtubules. The bleached area remained constant over time, consistent with 2D incompressible flows [Supp. Fig. 2]. The labeled nematic stretched along the director and compressed perpendicular to the director [Fig. \ref{FRAP}(a), Supp. Mov. 1]. Occasionally, the rotational component of the strain dominated, causing the region to shear [Supp. Fig. 3]. Our analysis excluded such regions. 

We defined the $x$-direction as the local nematic director in the Lagrangian reference frame. We first measured the intensity profile perpendicular to the director, $I_{\perp} (y)$ averaged across an edge of the photobleached region [Fig. \ref{FRAP}(b)]. The interface remained sharp while the width of the bleached region decayed exponentially as $w=ae^{-t/\tau}$, where $\gamma_{\perp}=1/\tau$ is the strain rate of the compressing material [Fig. \ref{FRAP}(c)]. The ATP dependent strain rate was described by the Michaelis-Menten equation, the same relationship also describing the ATP-dependent stepping of an isolated motor [Fig. \ref{FRAP}(g)]~\cite{SchnitzerForce}. 

Next, we measured the average intensity profiles along the director $I_{||}(x)$ averaged across the height of the bleached region. In contrast to the sharp edges perpendicular to the director, the edges along the director roughened [Fig. \ref{FRAP}(d)]. Over time, the average edge intensity broadened in a diffusion-like manner [Fig. \ref{FRAP}(e)]. We modeled the temporal evolution of such intensity profiles as a mass transfer process with diffusion and convection:
$\frac{\partial I_{||}}{\partial t}+\textbf{u}\cdot \nabla I_{||}=D_{\textrm{bleach}}\nabla^2 I_{||}$,
where $\textbf{u}$ is the velocity field and $D_{\textrm{bleach}}$ is the activity-induced effective diffusion coefficient. Restricting the flows to be dipolar extensile and incompressible gives: $\textbf{u}=\gamma_{||}\{x,-y\},$ where $\gamma_{||}$ is the constant strain rate. Assuming that gradients in the $y$ direction are negligible yields:
\begin{equation}\frac{\partial I_{||}}{\partial t}+x\gamma_{||}\frac{\partial I_{||}}{\partial x}=D_{\textrm{bleach}}\frac{\partial^2 I_{||}}{\partial x^2},\label{diffusion}\end{equation}
which we solved numerically. The strain rate and effective diffusion coefficients were obtained by minimizing the squared error between experimental intensity profiles and model predictions. The predicted intensity profiles quantitatively described the data [Fig. \ref{FRAP}(e)], yielding the ATP-dependent effective $D_{\textrm{bleach}}$ and $\gamma_{||}$ [Fig. \ref{FRAP}(f)]. Importantly, the measured strain rates were comparable to those extracted from the perpendicular intensity profiles, $I_{\perp} (y)$ [Fig. \ref{FRAP}(g)]. 

Photobleaching revealed the average dipolar extensile flows. Along the extension axis the bleached interface roughened, indicating that the microtubules slide past each other collectively in staggered bundles [Fig. \ref{FRAP}(d), Supp. Fig. 4]. This observation suggests the presence of complex spatiotemporal correlations in microtubule motions that remain poorly understood. Eq. \ref{diffusion} separates the dynamics into extensile motion that drives the large-scale chaotic flows ($\gamma_{||}$) and the stochastic contributions ($D_{\textrm{bleach}}$), by assuming that the microscopic motion of microtubules is the superposition of Gaussian random walk and spatially dependent drift processes. The similar ATP-dependent scaling of both $\gamma$ and $D_{\textrm{bleach}}$ suggests that both contributions are driven by the same microscopic processes.

\begin{figure}
\includegraphics[width=8.6cm]{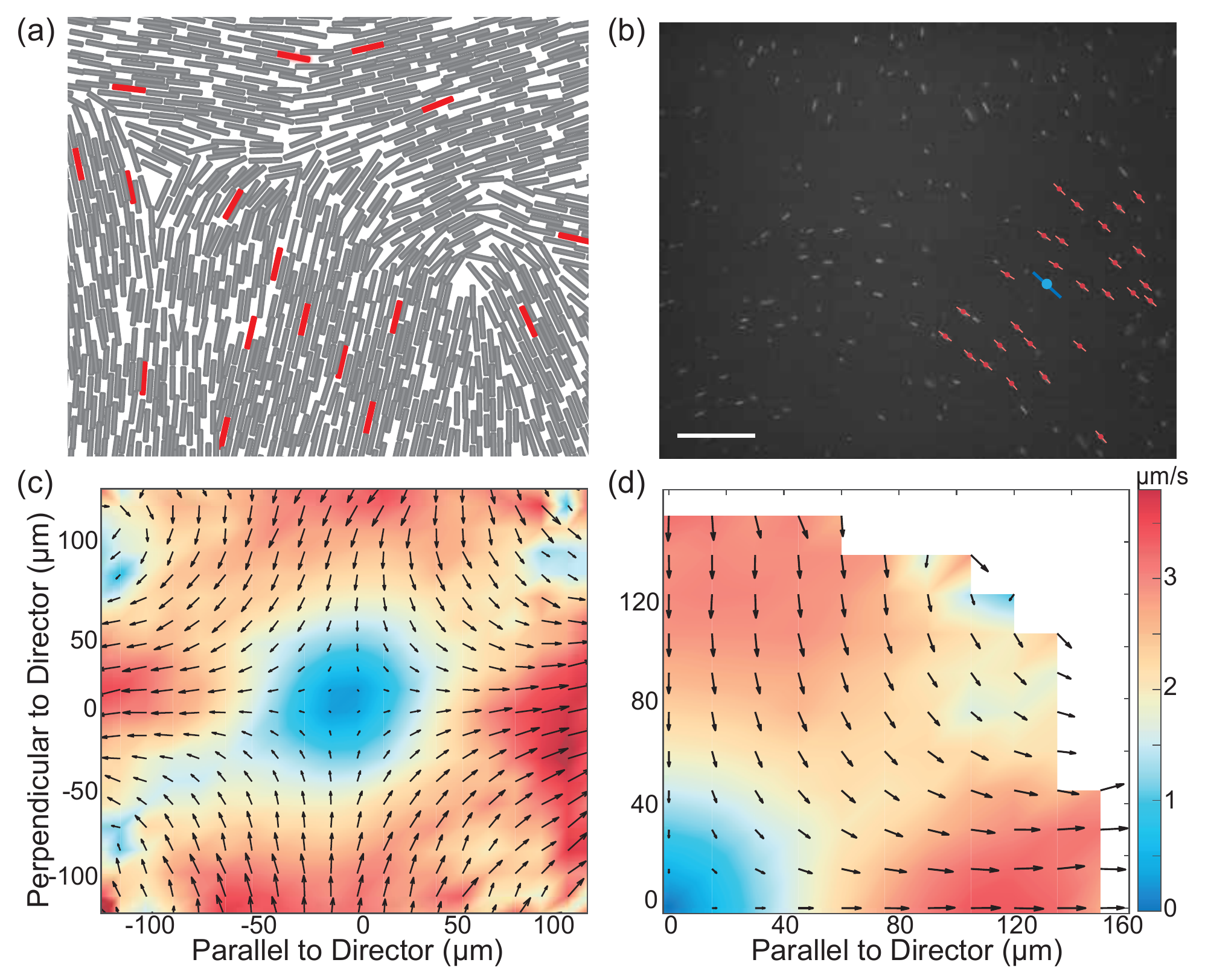}
\caption{\small \textbf{Tracking microtubules in active nematics.} \textbf{(a)} An active nematic doped with fluorescent microtubules. \textbf{(b)} An image of dilute fluorescent microtubules. The position (dot) and orientation (bar) of an example reference microtubule is overlaid in blue, and the aligned microtubules for which the relative velocities are calculated are plotted in red.  One out of twenty-thousand microtubules is labeled. Scale bar, 25 $\mu$m. \textbf{(c)} The average Lagrangian flow field in the reference frame of a microtubule located at the origin and oriented along $x$. The flow field was obtained from binning the relative velocities by distance and averaging over time and filament pairs (1000 $\mu$M ATP). \textbf{(d)} The average flow field with relative microtubule positions and velocities collapsed into the first quadrant, so that $V_{||}>0$ indicates an extending pair. }
\label{Filament Tracking Experiment}
\end{figure}

To probe the origin of the stochastic contributions, we quantified the relative motions of individual filaments in active nematics by labeling a low fraction of microtubules [Fig. \ref{Filament Tracking Experiment}(a)]. We extracted relative velocities and orientations of labeled filament pairs using a custom MATLAB program [Fig. \ref{Filament Tracking Experiment}(b), Supp. Mov. 2]. To analyze only uniform domains, the relative velocities were calculated only for filament pairs aligned to within $10\degree$. Averaging over all pairs yielded $V(x,y)$, which describes how the relative velocity of two microtubules depends on their separation along the $x$ and $y$ axis [Fig. \ref{Filament Tracking Experiment}(c)]. As previously, the microtubule's long axis defined the $x$-axis in the Lagrangian reference frame [Fig. \ref{Filament Tracking Experiment}(c)].  The velocities were broken into components along, $V_{||}(x,y)$, and perpendicular to, $V_{\perp}(x,y)$, the nematic director. Exploiting the nematic symmetry, we collapsed the flow field by rotating microtubule pairs so that the non-origin microtubule position was in the first quadrant [Fig. \ref{Filament Tracking Experiment}]. In this reference frame, $V_{||}>0$ indicates an extending pair, while $V_{||}<0$ indicates a contracting pair. Consistent with the photobleaching experiments, microtubule tracking also revealed dipolar extensile flows. 

\begin{figure}

\includegraphics[width=8.6cm]{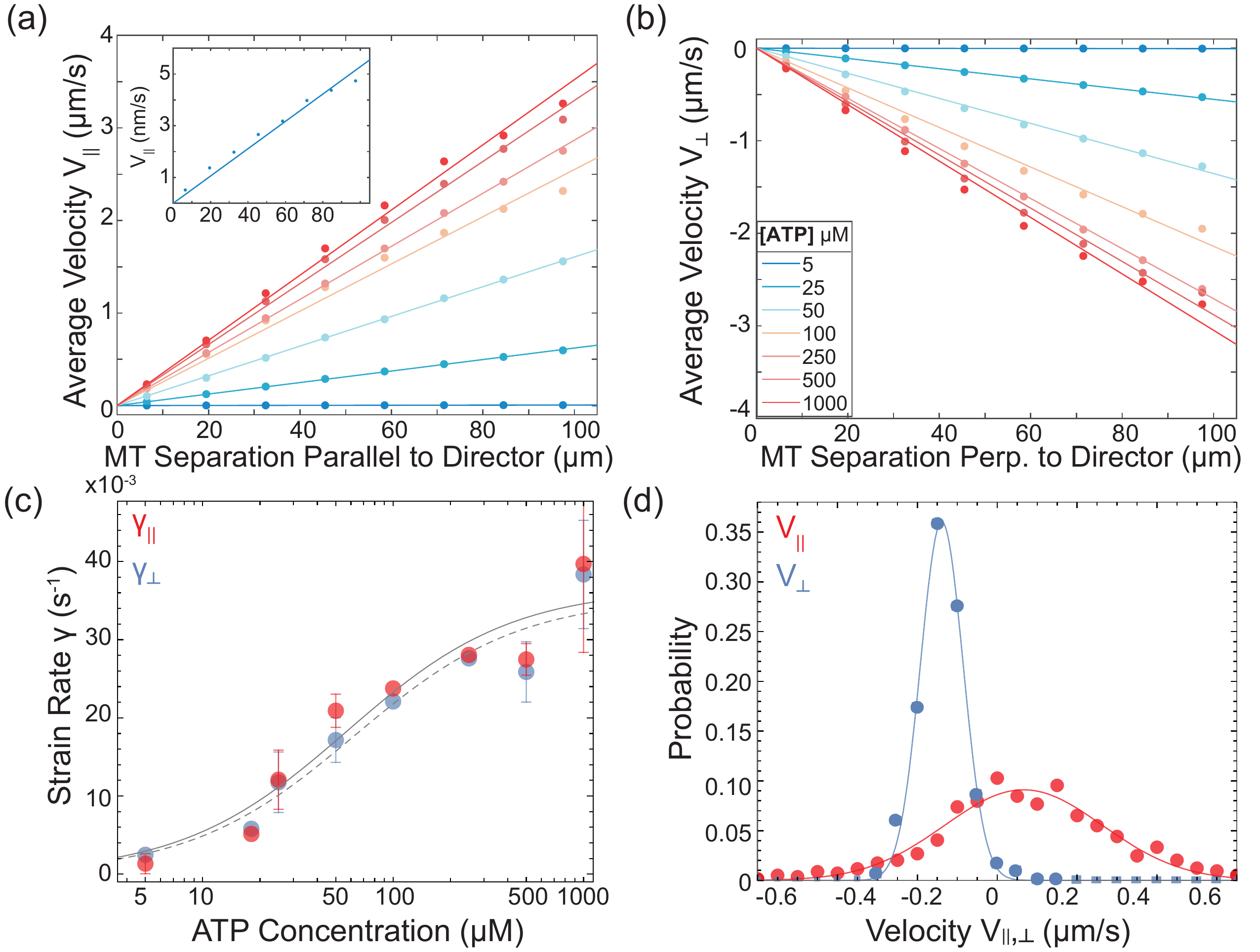}
\caption{\small \textbf{Analysis of filament tracking data.} \textbf{(a)} The average velocity of microtubule pairs along the director $V_{||}(x,0)$ as a function of interfilament distance along the director for all ATP concentrations [legend in panel (b)]. Inset: 5 $\mu$M velocity profile with a $y$-axis re-scaled to nm/s. \textbf{(b)} The average relative velocity perpendicular to the director $V_{\perp}(0,y)$ of microtubules as a function of the interfilament distance perpendicular to the director. \textbf{(c)} The strain rate calculated from the slope of the velocity profiles along the director (red) and perpendicular to the director (blue) plotted versus ATP concentration. The grey lines are fits of the data to the Michaelis-Menten equation with $\gamma_{||,\textrm{max}}=0.037$ $\textrm{s}^{-1}$ and $K_{||,\textrm{M}}=57$ $\mu$M (red, solid), and $\gamma_{\perp,\textrm{max}}=0.035$ $\textrm{s}^{-1}$ and $K_{\perp,\textrm{M}}=63$ $\mu$M (blue, dashed). Error bars are standard errors of multiple measurements. \textbf{(d) } The distribution of the microtubule sliding velocities for interfilament distance less than 6 $\mu$m (50 $\mu$M ATP). Lines are fits of the data to Gaussians $P=A*e^{[-(v-\bar{v})^2/2\sigma^2]}$. Along the director (red), $\bar{v}_{||}=0.04$ $\mu\textrm{m}/\textrm{s}$ and $\sigma_{||}=0.12$ $\mu\textrm{m}/\textrm{s}$. Perpendicular to the director (blue), $\bar{v}_{\perp}=-0.04$ $\mu$m/s and $\sigma_{\perp}=0.07$ $\mu$m/s. }
\label{Filament Tracking Results}

\end{figure}

\begin{figure}

\includegraphics[width=8.6cm]{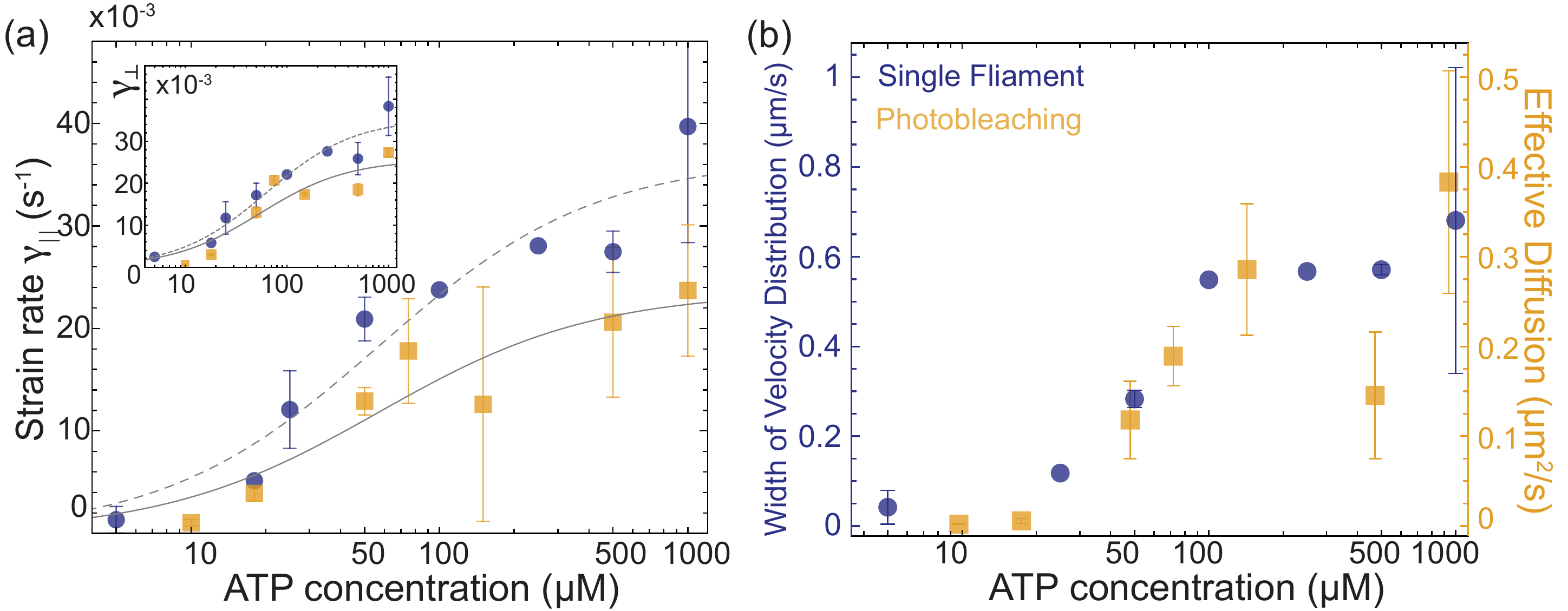}
\caption{\small \textbf{Comparison of photobleaching and filament tracking results.}  \textbf{(a) } Strain rate along the director extracted from the slope of velocity profiles (blue dots) and Eq. \ref{diffusion} (gold squares). Inset: Strain rate perpendicular to the director from slope of velocity profiles (blue dots) and exponential fit of photobleached region width (gold squares). Gray lines are fits to the Michaelis Menten equation.
\textbf{(b)} Width of velocity distribution along the director $\sigma_{||}$ (blue dots, left axis) and effective diffusion coefficient $D_{\textrm{bleach}}$ (gold squares, right axis) versus ATP concentration. }
\label{Comparison}

\end{figure}

Plotting the velocity profiles along the $x$ and $y$ axis, $V_{||}(x,0)$ and $V_{\perp}(0,y)$, yielded relative filament velocities that respectively increased/decreased linearly with increasing filament separation [Fig. \ref{Filament Tracking Results}(a-c)]. The slopes determined the ATP-dependent strain rates, which agreed with the photobleaching analysis [Fig. \ref{Comparison}(a)]. Both photobleaching and filament tracking analysis revealed constant average strain rate and net dipolar extensile flows in the defect-free regions of the active nematic. A constant strain rate indicates exponential extension along the director. Similar dynamics were observed in growing bacterial colonies, although in that case the influx of mass yielded isotropic growth \cite{Poon}.

Tracking individual filaments not only revealed the average extension rate, but also the distribution of velocities around the average value. The probability distributions of the microtubule separation velocities along, $p(V_{||})$, and perpendicular to, $p(V_{\perp})$, the director,  showed the deviations of individual filaments from the average strain rate [Fig. \ref{Filament Tracking Results}(d)]. Both distributions were Gaussians, with constant widths up to an interfilament distance of 30 $\mu$m [Supp. Fig. 5]. Perpendicular to the director, the relative velocities were tightly distributed: all microtubules moved toward each other at comparable speeds, corresponding to the sharp intensity profiles ($I_{\perp}$) of the photobleached region [Fig. \ref{FRAP}(a-b)]. By contrast, velocities along the director deviated significantly from the average strain rate. On average, microtubule pairs moved away from each other. However, $p(V_{||})$ had a significant contractile tail ($V_{||}<0$): at any given time  $37\%\pm2\%$ of microtubule pairs move toward each other. 

We extracted the width of the velocity distribution $\sigma_{||}$ from a Gaussian fit to $p(V_{||})$. Similar to the $D_{\textrm{bleach}}$ measured in the photobleaching experiments, $\sigma_{||}$ reflects the stochastic contribution of the microtubule movements. Indeed, we found that both variables scaled similarly with ATP concentration [Fig. \ref{Comparison}(b)]. For Brownian motion, the diffusion coefficient in the $x$ direction is related to the width of the velocity distribution $D_{\textrm{single}}= \sigma_{||}^2 \Delta t/2 $ \cite{Qian:1991}. However, we found that $D_{\textrm{single}}\neq D_{\textrm{bleach}}$ [Supp. Fig. 6]. The assumption of Brownian dynamics behind $D_{\textrm{single}}$ requires that particles act independently. In contrast, the jagged motion of microtubule bundles shows strong spatial correlations of microtubule motion [Fig. \ref{FRAP}(d)]. Taken together, these observations suggest that the underlying microscopic dynamics do not exhibit diffusion with Gaussian statistics.

Established models described active nematics by coarse-grained nematic director and velocity fields~\cite{LucaGeometry, VortexPaperLemma, RAlertActiveTurb,SurajDefectUnbinding, DefectsGiomi,DeCamp2015,NortonTheory, AchiniMike}. Such models capture certain features of the measured microtubule dynamics. Importantly, in a minimal model the local Lagrangian flows within a uniform domain exhibit relative extension speeds that grow linearly with distance [Supp. Fig. 7]. This agreement shows that the coarse-grained behaviors of active nematics are generic and may be obtained from distinct microscopic dynamics. However, our experiments show that the dynamics of individual microtubules which create the net extensional flows exhibit nontrivial deviations from the mean velocity. This is a first step for informing the development of multiscale theories. 

Our experiments revealed several features of active nematics. First, both stepping of isolated kinesin motors and the active nematic ATP-dependent strain rate are captured by the same Michaelis-Menten kinetic equations~\cite{SchnitzerForce}. This demonstrates that the average extension rate is determined by the kinesin stepping speed. Second, the strain rate of an isolated filament pair is estimated by dividing the extension velocity [Fig. \ref{Sliding}(b)] by the average microtubule length in the nematic, $\sim$2.5 $\mu$m. This microscopic strain rate ($\sim0.2 \textrm{s}^{-1}$) is significantly larger than the effective strain rate measured in active nematics (0.03 $\textrm{s}^{-1}$) where multi-motor and multi-filament interactions are present [Supp. Fig. 8]. This discrepancy suggests that the motors in dense nematics have different dynamics than those in the dilute solutions. One possibility is that motors in dense nematics experience high force loads, which reduces their stepping speed. Indeed, recent measurements suggested that the force load on kinesin is $\sim$20 pN which generates significant pre-stress~\cite{AlexPNAS2021}. Third, given the nearly uniform extension rate of isolated bundles, one might also expect uniform extension in the active nematic, wherein all microtubules slide past each other at the same rate \cite{GaoBetterton}. However, the measured distribution width of sliding velocities in dense nematics ($\sigma = 680$ nm/s) is much larger than what is found for dilute bundles ($\sigma = 100$ nm/s).  Finally, the narrow distribution of velocities perpendicular to the director $p(V_{\perp})$ is consistent with active stress being generated only along the director. The movement of microtubules perpendicular to the director is driven by incompressible material flows, with the contraction rate along $y$ reflecting the average extension rate along $x$. 

The microscopic origin of the extensile/contractile symmetry breaking remains an open question. Numerical and theoretical work suggests that non-uniform motor distributions along filaments could be the microscopic source of the symmetry breaking, but these distributions are challenging to measure \cite{Liverpool_2005, Belmonte2017, LenzeLife2020}. The measured filament velocity distributions provide an alternative parameter that can perhaps be used to confirm specific microscopic models. 

The microtubule velocity distributions determine the temporal evolution of a photobleached region. Uniform sliding velocities yield a photobleached area that splits into two distinct regions moving in the opposite directions, associated with opposing microtubule polarity \cite{GaoBetterton}. Such dynamics were  observed in 3D aligned microtubule gels powered by kinesin-14; all the microtubules moved at the same speed, exhibiting uniform extension that was identical to the isolated speed of the kinesin-14 motors \cite{BezSebastian}. In contrast, 2D active nematics exhibit non-uniform displacement along the edge of the bleached region. This striking difference suggests the existence of multiple classes of microscopic dynamics. Understanding the differences between these systems and their emergent dynamics is essential for developing accurate multiscale theories of active systems. 

Importantly, in our experiments the edge of the photobleached region along the director does not blur smoothly; rather it roughens as discrete ``bundles" of bleached and fluorescent microtubules slide past each other [Supp. Fig. 4, Supp. Mov. 3, 4]. This suggests that on short time scales bundles of microtubules move as cohesive units. One possibility is that the coherently moving regions have localized transient polar order~\cite{ShelleyReview, GaoBetterton}, a hypothesis which could be verified with second harmonic generation microscopy~\cite{Che-HangSHG, BezSebastian}. 

In conclusion, we probed the dynamics of microtubules within dense 2D active nematics. The average flows are dipolar extensile with well-defined average strain rates. The microscopic motions are widely distributed about this average and cannot be captured by coarse-grained variables. The velocity distribution along the director and the appearance of coherent bundle motion implies that non-additive motor-microtubule interactions are important for generating extensile active stress. 

\acknowledgments{\textbf{Acknowledgments:} This manuscript was primarily supported by the Department of Energy Basic Energy Science  through award DE-SC0019733 (L.M.L., S.J.D and Z.D.). M.F.H. and S.F. acknowledges support from NSF DMR-1855914 and the Brandeis Center for Bioinspired Soft Materials, an NSF MRSEC (DMR-1420382 and DMR-2011846). A.M.T. is a Simons Foundation Fellow of the Life Sciences Research Foundation and is an Awardee of the Weizmann Institute of Science–National Postdoctoral Award Program for Advancing Women in Science. We also acknowledge computational support from NSF XSEDE computing resources allocation TG-MCB090163 (Stampede) and the Brandeis HPCC which is partially supported by DMR-1420382. We also acknowledge the KITP Active20 program, which is supported in part by the National Science Foundation under Grant No. NSF PHY-1748958. }

\clearpage
\onecolumngrid
\begin{center}
 \textbf{\large{Supplementary Information for Multiscale Microtubule Dynamics in Active Nematics}}
August 9, 2021
\end{center}

\section{Methods}

\subsection{Active Nematic Preparation}

The active material was composed of microtubules, kinesin motor clusters, a depletion agent and an ATP regeneration system. The dynamics were sensitive to the purification of proteins and preparation of the chemical stocks. To ensure quantitatively reproducible results, a large-scale active mixture was prepared and stored at $-80\degree$C in individual aliquots until the day of experiment. All the results in the manuscript originated from the same stock. 

We prepared the mixture following the previously published protocol \cite{Sanchez2012, DeCamp2015}. 
Briefly, a truncated version of Kinesin-1 with a biotin-accepting domain (K401-BCCP-HIS) was expressed in the presence of biotin from Rosetta2(DE3) pLysS \textit{E. Coli} and purified using immobilized metal affinity chromatography \cite{Subramanian2007}. Biotinylated kinesin (5 $\mu$L of 0.7 mg/mL) was mixed with streptavidin (5.7 $\mu$L of 0.35 mg/mL) and incubated for 30 min to assemble motor clusters. The active mixture was composed of polyethylene glycol (0.8\% w/v 20 kDa, Sigma Aldrich) as a depletion agent, phosphoenol pyruvate (26.6 mM, Beantown Chemicals) and pyruvate kinase/lactic dehydrogenase (PK/LDH Sigma, P-0294) as ATP regeneration system, and  glucose (6.7 mg/mL), glucose catalase (0.4 mg/mL), glucose oxidase (0.08 mg/mL) and trolox (2 mM) as an anti-oxidant system. The components were mixed in M2B buffer (80 mM PIPES, pH 6.8, 1 mM EGTA, 2 mM MgCl$_2$) and the MgCl$_2$ concentration was brought up to 5 mM. ATP (Sigma Aldrich) was added in the desired concentration (from 5 $\mu$M to 1000 $\mu$M) to individual aliquots of the active mixture just before flash freezing in liquid nitrogen.

The tubulin was purified from bovine brain \cite{PopovTubulin}. The tubulin was either labeled with NHS-Alexa 647 (Invitrogen) or recycled without labeling \cite{HymanTubulin}. The tubulin was polymerized at 8 mg/mL in the presence of Guanosine-5-[($\alpha,\beta$)-methyleno]triphosphate GMPCPP (Jena Bioscience NU-405L) to stabilize the microtubules to an average length of 2.5 $\mu$m. Two batches of microtubules were polymerized: one of only unlabeled tubulin and one with a final fraction of 3\% labeled tubulin. Each polymerization process can result in different length distribution, so using microtubules from the same polymerization is important to keep this variable from affecting measurements. 

We used a flow chamber with dimensions 18 $\times$ 3 $\times$ 0.06 mm made of laser cut double-sided tape sandwiched between a glass slide and coverslip. We treated the bottom slide with commercially available aquapel to make the surface hydrophobic, and the coverslip was passivated with acrylamide \cite{Lau_2009}. These chambers hold 7 $\mu$L of sample. To form a large, flat 2D interface, first we flowed in oil (HFE 7500) stabilized with a fluoro-surfactant PFPE–PEG–PFPE (1.8\% w/v, RAN Biotech), and then flowed in the aqueous active mixture while wicking out the oil. This left a thin 2D layer of surfactant stabilized oil-water interface next to the hydrophobic surface. Over time extensile microtubule bundles sedimented to the oil-water interface. The chamber was sealed with Norland Optical Adhesive, cured under a UV lamp for 1 min. The sedimentation of microtubules onto the interface was aided by centrifuging the sample in a swinging bucket centrifuge (Sorval Legend RT rotor 6434) at 1000 RPM for 20 min. 

On the day of the experiment, the pre-mixed active components and microtubules were rapidly thawed. Unlabeled and/or labeled microtubules were added and loaded into the chamber as described. 

\newpage

\subsection{Dilute Microtubule Extension Preparation}

3\% Alexa 647 microtubules were polymerized at 0.2 mg/mL in the presence of Guanosine-5-[($\alpha,\beta$)-methyleno]triphosphate GMPCPP (Jena Bioscience NU-405L) to stabilize the microtubules to $\sim$ 10 $\mu$m. These long microtubules were added to the same active mixture as the active nematics with 1.4 mM ATP for a final tubulin concentration of $10^{-3}$ mg/mL. Long microtubules were used to increase the statistics for the extension rate. Since the extension rate is constant with filament overlap, the length of the microtubule should not affect this measurement. As discussed in the main text, any contribution due to drag of the filament is very small compared to the stall forces of the motors.  

We used the same flow chambers as for the dense active nematics and prepared the sample identically. Samples were imaged with a Nikon Ti-Eclipse microscope, 60X Plan Fluor Oil (1.25 NA) and Andor Zyla camera. 

\subsection{Photobleaching}

A Nikon Ti-Eclipse microscope was customized to make the multiphoton excitation and wide field illumination simultaneously possible. A femtosecond pulse laser (Mai Tai HP DeepSee Sapphire) with DCS-120 scanning system and a HPM-100 detector (Becker and Hickl GmbH) required for photobleaching were installed on one optical arm of the microscope. A widefield illumination and sCMOS camera (Andor Zyla) were installed on the second arm for fluorescent imaging. The peak power of the laser was 56 kW at 690 nm, the excitation wavelength of the microtubules. However, this power causes cavitation within in the sample and destroys microtubules. An AOM synchronized with scanning system was used to adjust the average laser power to 30-75 mW which photobleached the sample quickly enough to limit distortions due to activity, while also minimizing damage to the proteins. In control experiments, we used the second harmonic generation of microtubules to confirm that microtubules were only bleached and not ablated at the laser powers used \cite{Che-HangSHG}.  We observed the deformation of the square using the fluorescence arm of the Ti-Eclipse. Intensity profiles were measured using ImageJ, averaging across the length of the edge. 

\subsection{Diffusion-Convection Model}
All experimental intensity profiles were normalized by the average intensity in a region far from the bleach. Eq. 1 was numerically solved using the method of lines. The simulations were initialized with data directly from the experiments. Dirichlet boundary conditions corresponding to the minimum and maximum intensity are applied at the left and right boundaries, respectively, but are sufficiently far from the interface so as not to impact the solution. Strain rate and diffusion coefficients were determined by minimizing the squared error between experimental intensity profiles and model predictions using MATLAB's lsqnonlin() function.

\subsection{Filament Tracking}

An inverted Nikon Ti-Eclipse with an Andor Neo camera was used to observe the samples using epi-fluorescence microscopy. The large field of view of the Neo camera yielded sufficient statistics of microtubule-pairs at large inter-filament distance. A 100X 1.3NA objective (theoretical resolution of 0.3 $\mu$m) was used in order to resolve both the position and orientation of microtubules. The microtubule identification and tracking was done using a custom MATLAB program which used regionprops() to find microtubules and fit them to ellipses. The properties and positions of the ellipses were used to identify filaments and track them through time. 

Relative positions and velocities were calculated from microtubule trajectories. Since the system has nematic symmetry, we defined our frame of reference to always have one filament position at the origin and the other filament position in the first quadrant $[+x,+y]$ so that negative velocities $V_{||,\perp}<0$ indicated a contracting pair and positive velocities $V_{||,\perp}>0$ indicated an extending pair. Velocities were binned by position with bin size of 6 $\mu$m along both $x$ and $y$ [Supp. Fig. \ref{Vx_Distribution}(a)]. The first bin has a finite lower limit due to the inability to resolve overlapping microtubules. Velocity profiles were obtained by taking $V_{||}(x,0)$ ($V_{\perp}(0,y)$) where $x=0$ ($y=0$) indicates the first bin of data. 

To obtain velocity distributions, we first set a threshold for the $y$ separation of microtubules, binsize$_y$=6 $\mu$m [Supp. Fig. \ref{Vx_Distribution}(a)]. Plotting the $x$-component of the relative velocities versus the $x$-separation yields a scatter plot [Supp. Fig. \ref{Vx_Distribution}(b)]. These velocities are then binned along $x$ and a width of the velocity distribution $\sigma$ is extracted from Gaussian fit to the probability distribution from each bin [Supp. Fig. \ref{Vx_Distribution}(c)]. The width $\sigma$ is relatively constant for increasing $x$-separation, until the number of microtubule pairs becomes small [Supp. Fig. \ref{Vx_Distribution}(d)]. The bin size was optimized to minimize the width of the velocity distributions [Supp. Fig. \ref{Vx_Distribution}(d), inset]. The distributions of $V_{\perp}$ along $y$ are obtained through the same method and behave similarly.  

\section{Estimate of Drag Forces Between Microtubules}

From \cite{Ward2015}, the diffusion coefficient between two microtubules is 0.14 $\mu\textrm{m}^2$/s which leads to a friction coefficient of $f =3\times10^{-5}$ pN s $\textrm{nm}^{-1}$. Therefore the frictional drag force experienced by the microtubule pairs extending at 540 nm/s is $F=fv\sim 10^{-2}$ pN. Besides a frictional drag force, the microtubules also experience a compaction force due to the depletion agent in the system estimated to be 25 k$_\textrm{B}$T \cite{FeodorFriction}. At room temperature this force is $\sim 10^{-1}$ pN. Finally the drag force due to pulling a rod (25 nm diameter, 10 $\mu$m length) through a fluid with the viscosity of water is $\sim 10^{-4}$ pN. All combined the drag forces due to the load on the kinesin clusters is negligible in comparison to the stall force of the motors, indicating that the force angle and oil-water interface are the dominating explanations for the slow speed of extension compared to that expected from single molecule experiments. 

\section{Continuum Theory of Active Nematic}

A continuum model that captures the basic features of an active nematic is governed by the fluid equation presented
in dimensionless form: 
$\nabla^2 u -\nabla P - \alpha \nabla\cdot Q=0,$
 where
$u$ is the velocity field, $P$ is a pressure field, $\alpha$ is the activity parameter, and $Q$ is the nematic order parameter
\cite{AchiniMike, NortonTheory}. For simplicity, we assume that viscous and active
stresses dominate the force balance and ignore passive
elastic stresses. This model exhibits $\pm$1/2 defects that
are characteristic of the extensile active nematic [Supp. Fig. \ref{Continuum}(a)]. Within an aligned region, the local flow is extensile
with the principle axis along the local director field [Supp. Fig. \ref{Continuum}(b)]. The relative velocity between two points in the
nematic monodomain grows linearly with distance, up
to the length scale where the nematic order is spatially
uniform, in agreement with experiments [Supp. Fig. \ref{Continuum}(c)]. In
the model, aligned regions have uniform macroscopic active stress $\alpha Q$, which means there is no flow-generating
momentum flux. This suggests that the flow-generating
forces arise entirely at the boundary of uniform regions.

 \section{Movies}
\noindent \textbf{Supplementary Movie 1:} Evolution of a photobleached region in an active nematic (18 $\mu$M ATP).  
\\
\\
\noindent \textbf{Supplementary Video 2:} Left: Fluorescent microscopy movie of a nematic with a low fraction labeled microtubules (50 $\mu$M ATP). Right: The positions and orientations of microtubules as measured by custom MATLAB program. Color identifies individual microtubules whose trajectories are tracked.
\\
\\
\noindent \textbf{Supplementary Movie 3 \& 4:} High magnification movies of time evolution of photobleached region in active nematic (10 $\mu$M ATP).

\newpage

\section{Figures}

\begin{figure}[hbt!]

\includegraphics[width=12cm]{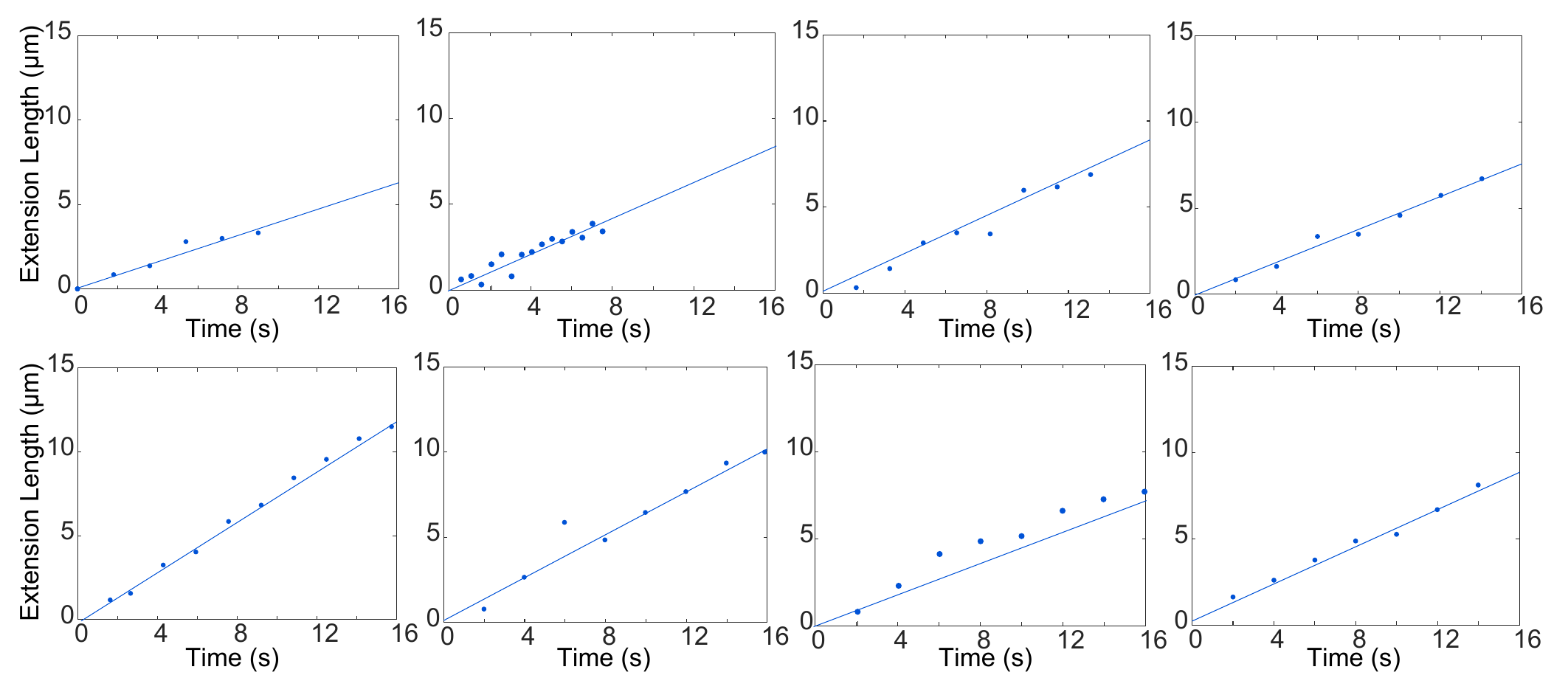}
\caption{\small Bundle extension length versus time for eight microtubule pairs. Each extension is fit to a line, the slope of which reflects the extension speed of the bundle. }
\end{figure}

\begin{figure} [hbt!]

\includegraphics[width=12cm]{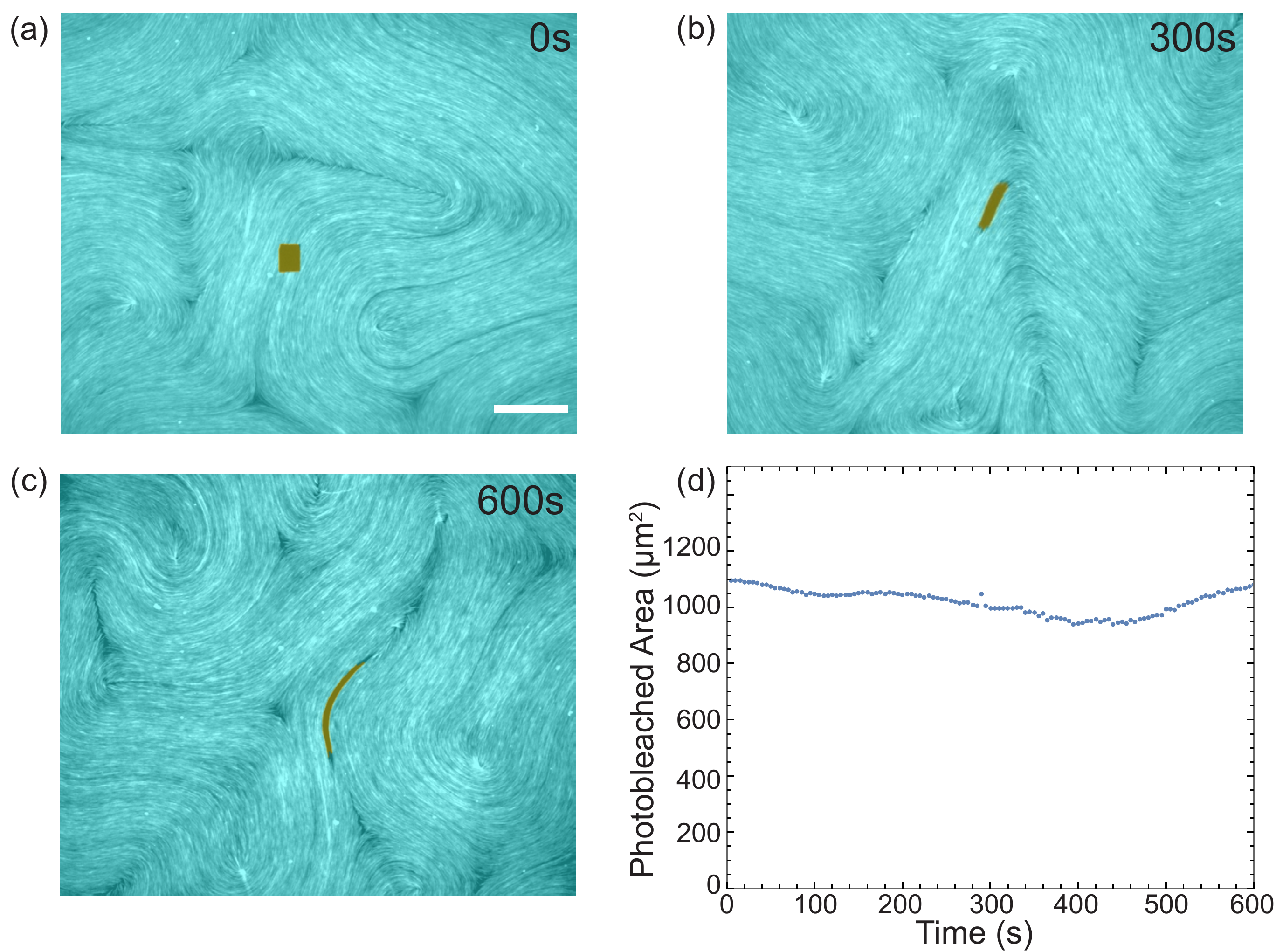}
\caption{\small  \textbf{(a)-(c)} Time series of a photobleached region (18 $\mu$M ATP). The color indicates the regions identified by the pixel classifier ilastik as within the bleach region (gold) and outside the bleached region (cyan) \cite{ilastik}. Scale bar 100 $\mu$m. \textbf{(d)} The area of the identified photobleached region versus time. }

\end{figure}

\begin{figure} 

\includegraphics[width=12cm]{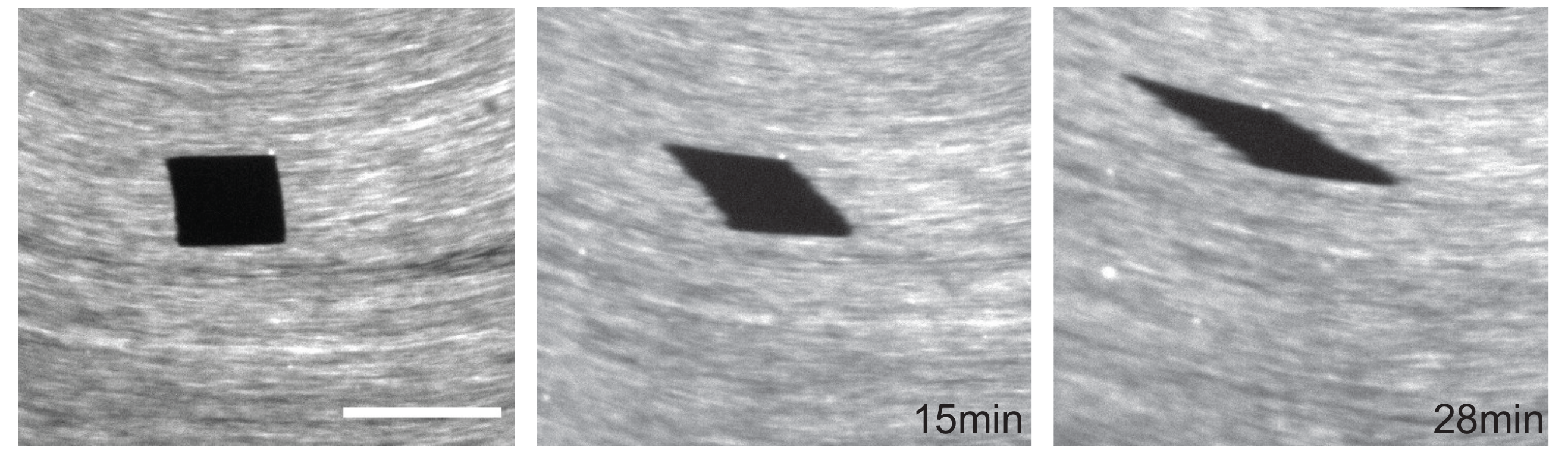}
\caption{\small  Time series of a photobleached region aligned with nematic director in a region where shear dominates (10 $\mu$M ATP). Scale bar 50 $\mu$m. }

\end{figure}

 \begin{figure}

\includegraphics[width=9cm]{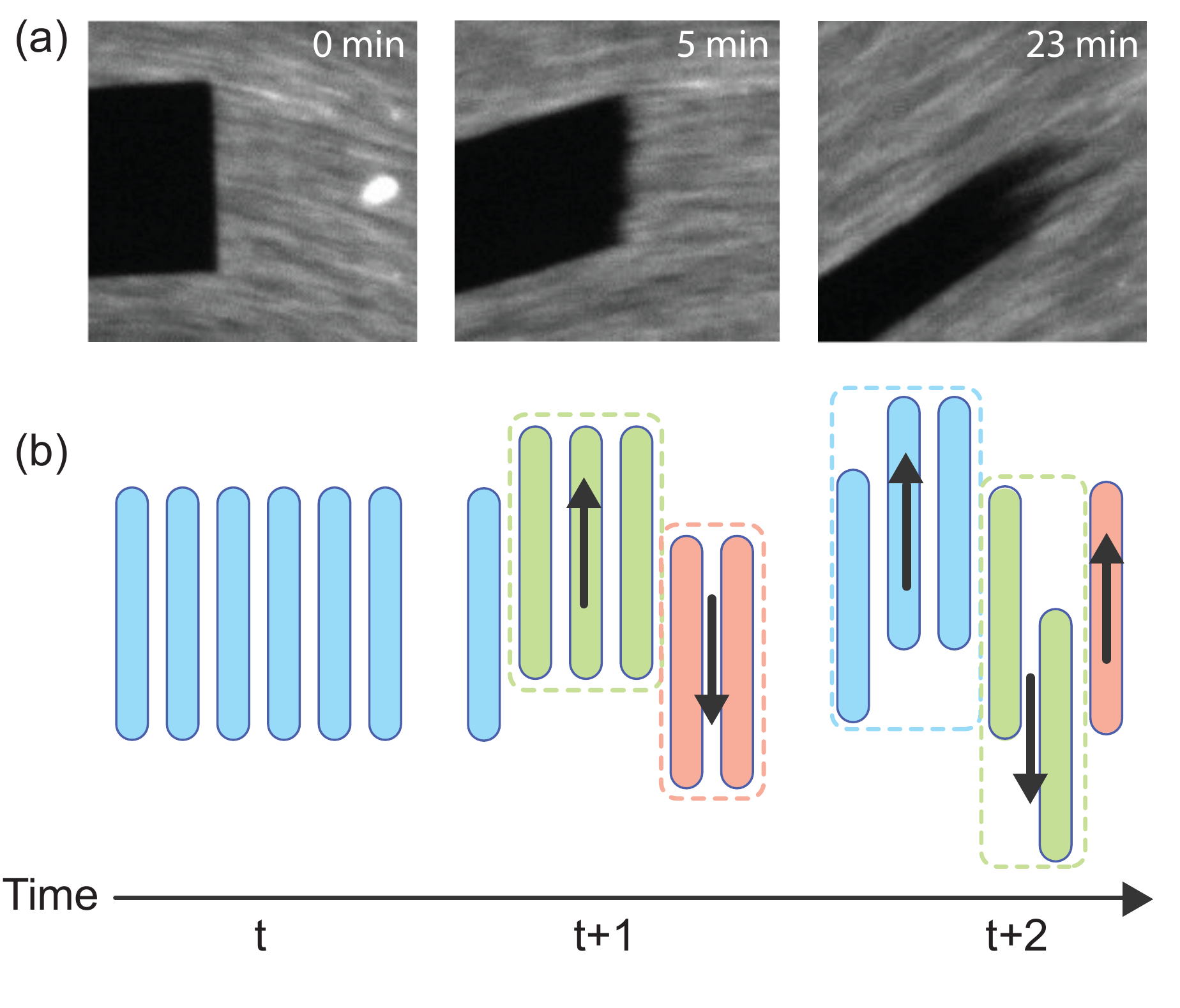}
\caption{\small  \textbf{(a)} Experimental time series of sharp photobleached edge roughening as fluorescent microtubules mix with the bleached. \textbf{(b)} A schematic of the non-uniform sliding of microtubules which causes the observed roughening.  }

\end{figure}

 \begin{figure}

\includegraphics[width=16cm]{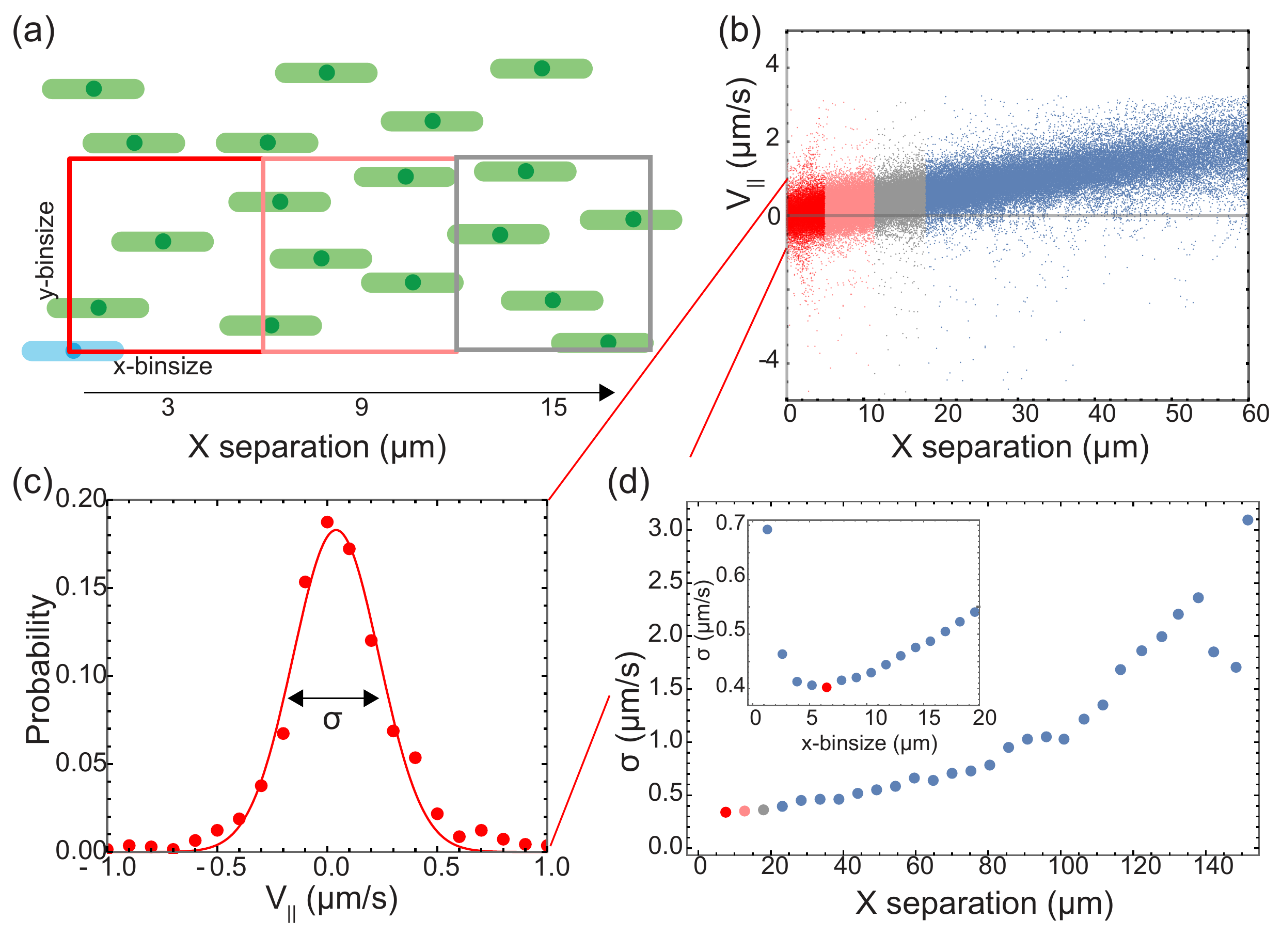}
\caption{\small  \textbf{(a)} The origin microtubule (blue) with velocity pairs with positions in the first quadrant (green). The boxes indicate the bin sizes along ($x$) and perpendicular to ($y$) the director. \textbf{(b)} All relative velocities whose $y$-separation is less than $y$-binsize = 6$\mu$m plotted as a function of their $x$-separation. \textbf{(c)} Histogram of the velocities in the first bin. The line indicates a fit to a Gaussian with standard deviation $\sigma$. We take $\sigma$ to describe the width of the distribution. \textbf{(d)} Width of velocity distribution for bins at different $x$-separation. Inset: Width of velocity distribution for different bin sizes along $x$.}

\label{Vx_Distribution}
\end{figure}

 \begin{figure}

\includegraphics[width=9cm]{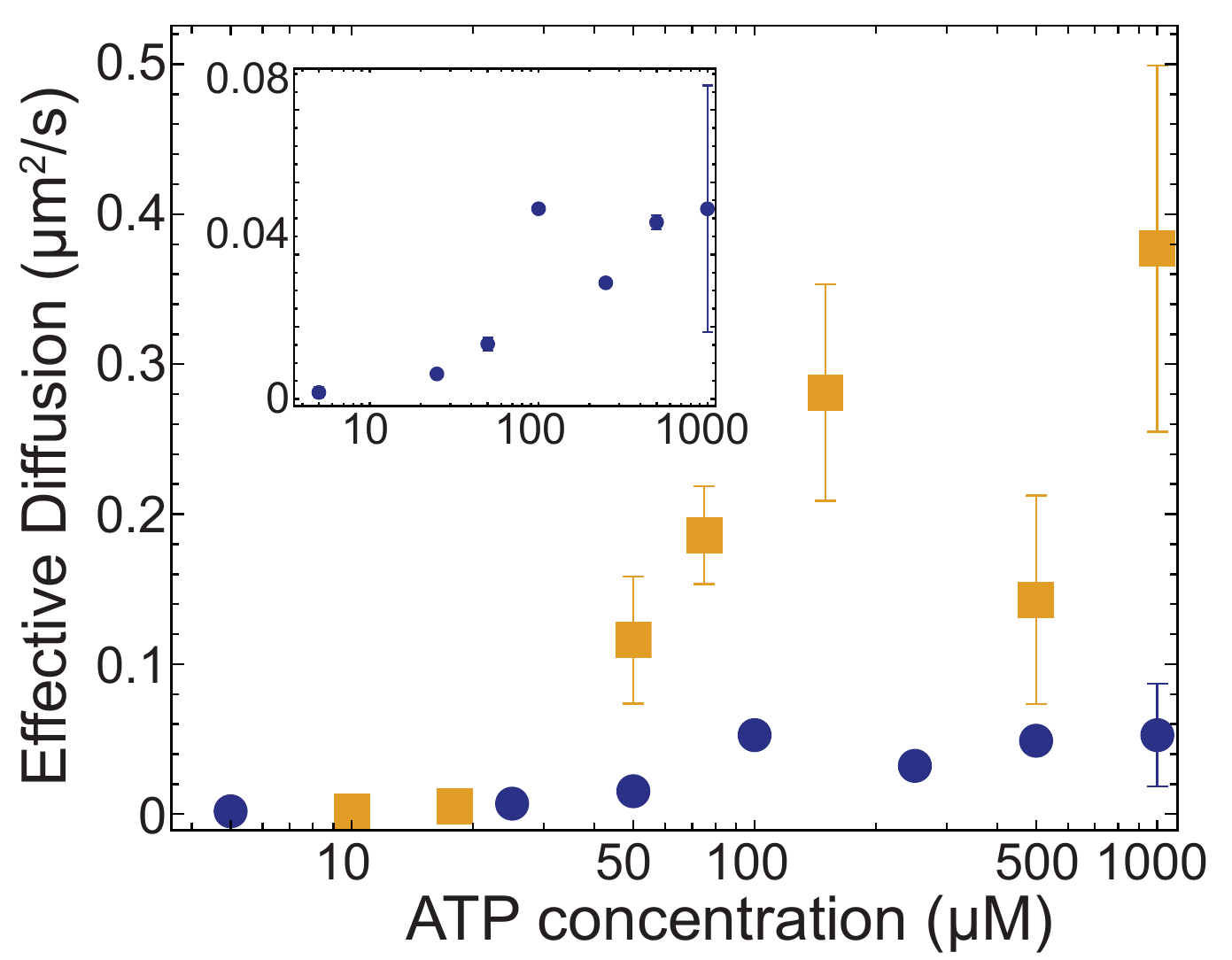}
\caption{\small  Effective diffusion coefficient extracted from fit to Eq. 1 with photobleaching data (yellow squares) and from single filament velocity distributions $D_{\textrm{single}}=\sigma_{||}^2\Delta t/2$ (blue circles) as a function of ATP concentration in active nematic. Inset: re-scaled $y$-axis for $D_{\textrm{single}}$.}

\end{figure}

\begin{figure}

\includegraphics[width=12cm]{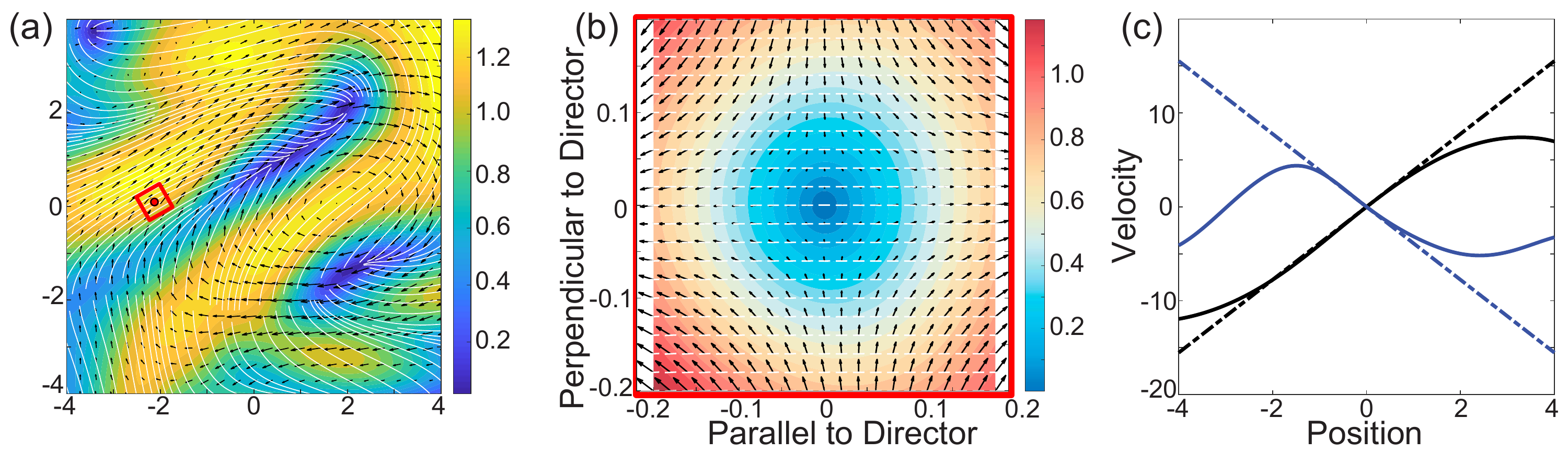}
\caption{\small  Flows in a continuum model of active nematics. \textbf{(a)}  The global flow field (black arrows) of a bulk active
nematic. White lines indicate the director field and color
maps to the order parameter. \textbf{(b)} The local flow in the red
square from (a) in the reference frame of the fluid. The color
indicates the relative speed of the flows. \textbf{(c)} The velocity
profile along (black) and perpendicular (blue) to the director.
The dotted lines are linear fits to the initial portion. The
scales are all in dimensionless units. }

\label{Continuum}
\end{figure}

\begin{figure}

\includegraphics[width=9cm]{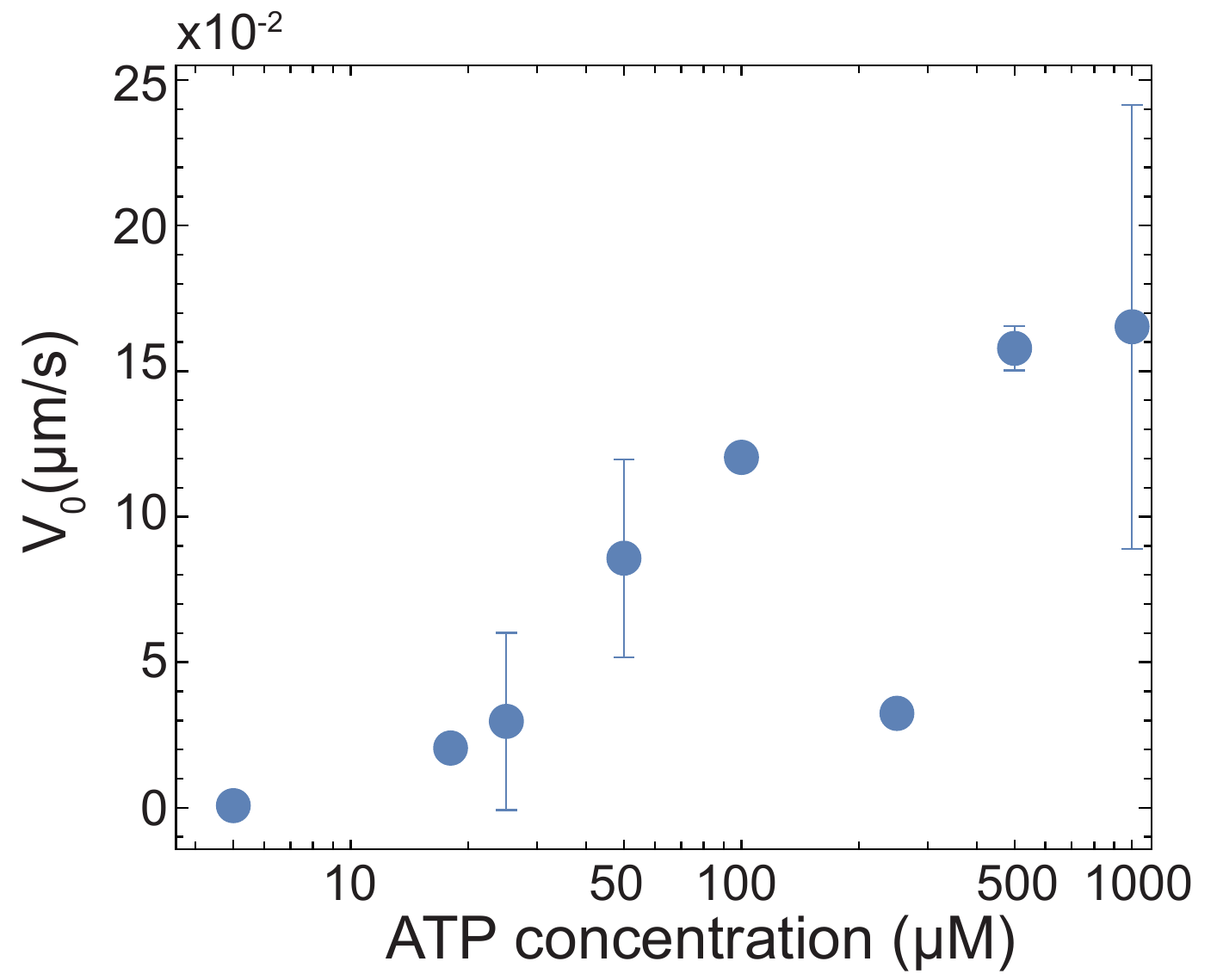}
\caption{\small  Instead of looking at strain rate, one can also compare the extension velocity at zero microtubule separation in the active nematic to the extension velocity of dilute microtubule pairs (540 nm/s). Here, in the dense active nematic the extrapolated velocity at 0 $\mu$m inter-filament separation along the director flow axis ($V_0$) versus ATP concentration. $V_0$ is estimated as the $y$-intercept of linear fits to velocity profiles along the director [Fig. 3(a)]. Error bars indicate the standard error of multiple measurements. At saturating ATP, the average extension velocity is 150 nm/s, significantly lower than in the dilute regime.}

\end{figure}


\begin{thebibliography}{56}%
\makeatletter
\providecommand \@ifxundefined [1]{%
 \@ifx{#1\undefined}
}%
\providecommand \@ifnum [1]{%
 \ifnum #1\expandafter \@firstoftwo
 \else \expandafter \@secondoftwo
 \fi
}%
\providecommand \@ifx [1]{%
 \ifx #1\expandafter \@firstoftwo
 \else \expandafter \@secondoftwo
 \fi
}%
\providecommand \natexlab [1]{#1}%
\providecommand \enquote  [1]{``#1''}%
\providecommand \bibnamefont  [1]{#1}%
\providecommand \bibfnamefont [1]{#1}%
\providecommand \citenamefont [1]{#1}%
\providecommand \href@noop [0]{\@secondoftwo}%
\providecommand \href [0]{\begingroup \@sanitize@url \@href}%
\providecommand \@href[1]{\@@startlink{#1}\@@href}%
\providecommand \@@href[1]{\endgroup#1\@@endlink}%
\providecommand \@sanitize@url [0]{\catcode `\\12\catcode `\$12\catcode
  `\&12\catcode `\#12\catcode `\^12\catcode `\_12\catcode `\%12\relax}%
\providecommand \@@startlink[1]{}%
\providecommand \@@endlink[0]{}%
\providecommand \url  [0]{\begingroup\@sanitize@url \@url }%
\providecommand \@url [1]{\endgroup\@href {#1}{\urlprefix }}%
\providecommand \urlprefix  [0]{URL }%
\providecommand \Eprint [0]{\href }%
\providecommand \doibase [0]{https://doi.org/}%
\providecommand \selectlanguage [0]{\@gobble}%
\providecommand \bibinfo  [0]{\@secondoftwo}%
\providecommand \bibfield  [0]{\@secondoftwo}%
\providecommand \translation [1]{[#1]}%
\providecommand \BibitemOpen [0]{}%
\providecommand \bibitemStop [0]{}%
\providecommand \bibitemNoStop [0]{.\EOS\space}%
\providecommand \EOS [0]{\spacefactor3000\relax}%
\providecommand \BibitemShut  [1]{\csname bibitem#1\endcsname}%
\let\auto@bib@innerbib\@empty
\bibitem [{\citenamefont {Marchetti}\ \emph {et~al.}(2013)\citenamefont
  {Marchetti}, \citenamefont {Joanny}, \citenamefont {Ramaswamy}, \citenamefont
  {Liverpool}, \citenamefont {Prost}, \citenamefont {Rao},\ and\ \citenamefont
  {Simha}}]{MarchettiReview}%
  \BibitemOpen
  \bibfield  {author} {\bibinfo {author} {\bibfnamefont {M.~C.}\ \bibnamefont
  {Marchetti}}, \bibinfo {author} {\bibfnamefont {J.~F.}\ \bibnamefont
  {Joanny}}, \bibinfo {author} {\bibfnamefont {S.}~\bibnamefont {Ramaswamy}},
  \bibinfo {author} {\bibfnamefont {T.~B.}\ \bibnamefont {Liverpool}}, \bibinfo
  {author} {\bibfnamefont {J.}~\bibnamefont {Prost}}, \bibinfo {author}
  {\bibfnamefont {M.}~\bibnamefont {Rao}},\ and\ \bibinfo {author}
  {\bibfnamefont {R.~A.}\ \bibnamefont {Simha}},\ }\bibfield  {title} {\bibinfo
  {title} {Hydrodynamics of soft active matter},\ }\href@noop {} {\bibfield
  {journal} {\bibinfo  {journal} {Rev. Mod. Phys.}\ }\textbf {\bibinfo {volume}
  {85}},\ \bibinfo {pages} {1143} (\bibinfo {year} {2013})}\BibitemShut
  {NoStop}%
\bibitem [{\citenamefont {Narayan}\ \emph {et~al.}(2007)\citenamefont
  {Narayan}, \citenamefont {Ramaswamy},\ and\ \citenamefont {Menon}}]{Narayan}%
  \BibitemOpen
  \bibfield  {author} {\bibinfo {author} {\bibfnamefont {V.}~\bibnamefont
  {Narayan}}, \bibinfo {author} {\bibfnamefont {S.}~\bibnamefont {Ramaswamy}},\
  and\ \bibinfo {author} {\bibfnamefont {N.}~\bibnamefont {Menon}},\ }\bibfield
   {title} {\bibinfo {title} {Long-lived giant number fluctuations in a
  swarming granular nematic},\ }\href@noop {} {\bibfield  {journal} {\bibinfo
  {journal} {Science}\ }\textbf {\bibinfo {volume} {317}},\ \bibinfo {pages}
  {105} (\bibinfo {year} {2007})}\BibitemShut {NoStop}%
\bibitem [{\citenamefont {Zhou}\ \emph {et~al.}(2014)\citenamefont {Zhou},
  \citenamefont {Sokolov}, \citenamefont {Lavrentovich},\ and\ \citenamefont
  {Aranson}}]{LLCIgor}%
  \BibitemOpen
  \bibfield  {author} {\bibinfo {author} {\bibfnamefont {S.}~\bibnamefont
  {Zhou}}, \bibinfo {author} {\bibfnamefont {A.}~\bibnamefont {Sokolov}},
  \bibinfo {author} {\bibfnamefont {O.~D.}\ \bibnamefont {Lavrentovich}},\ and\
  \bibinfo {author} {\bibfnamefont {I.~S.}\ \bibnamefont {Aranson}},\
  }\bibfield  {title} {\bibinfo {title} {Living liquid crystals},\ }\href@noop
  {} {\bibfield  {journal} {\bibinfo  {journal} {Proceedings of the National
  Academy of Sciences}\ }\textbf {\bibinfo {volume} {111}},\ \bibinfo {pages}
  {1265} (\bibinfo {year} {2014})}\BibitemShut {NoStop}%
\bibitem [{\citenamefont {Saw}\ \emph {et~al.}(2017)\citenamefont {Saw},
  \citenamefont {Doostmohammadi}, \citenamefont {Nier}, \citenamefont
  {Kocgozlu}, \citenamefont {Thampi}, \citenamefont {Toyama}, \citenamefont
  {Marcq}, \citenamefont {Lim}, \citenamefont {Yeomans},\ and\ \citenamefont
  {Ladoux}}]{EpitheliaNematic}%
  \BibitemOpen
  \bibfield  {author} {\bibinfo {author} {\bibfnamefont {T.~B.}\ \bibnamefont
  {Saw}}, \bibinfo {author} {\bibfnamefont {A.}~\bibnamefont {Doostmohammadi}},
  \bibinfo {author} {\bibfnamefont {V.}~\bibnamefont {Nier}}, \bibinfo {author}
  {\bibfnamefont {L.}~\bibnamefont {Kocgozlu}}, \bibinfo {author}
  {\bibfnamefont {S.}~\bibnamefont {Thampi}}, \bibinfo {author} {\bibfnamefont
  {Y.}~\bibnamefont {Toyama}}, \bibinfo {author} {\bibfnamefont
  {P.}~\bibnamefont {Marcq}}, \bibinfo {author} {\bibfnamefont {C.~T.}\
  \bibnamefont {Lim}}, \bibinfo {author} {\bibfnamefont {J.~M.}\ \bibnamefont
  {Yeomans}},\ and\ \bibinfo {author} {\bibfnamefont {B.}~\bibnamefont
  {Ladoux}},\ }\bibfield  {title} {\bibinfo {title} {Topological defects in
  epithelia govern cell death and extrusion},\ }\href
  {https://doi.org/10.1038/nature21718} {\bibfield  {journal} {\bibinfo
  {journal} {Nature}\ }\textbf {\bibinfo {volume} {544}},\ \bibinfo {pages}
  {212} (\bibinfo {year} {2017})}\BibitemShut {NoStop}%
\bibitem [{\citenamefont {Kawaguchi}\ \emph {et~al.}(2017)\citenamefont
  {Kawaguchi}, \citenamefont {Kageyama},\ and\ \citenamefont
  {Sano}}]{SanoNematic}%
  \BibitemOpen
  \bibfield  {author} {\bibinfo {author} {\bibfnamefont {K.}~\bibnamefont
  {Kawaguchi}}, \bibinfo {author} {\bibfnamefont {R.}~\bibnamefont
  {Kageyama}},\ and\ \bibinfo {author} {\bibfnamefont {M.}~\bibnamefont
  {Sano}},\ }\bibfield  {title} {\bibinfo {title} {Topological defects control
  collective dynamics in neural progenitor cell cultures},\ }\href
  {https://doi.org/10.1038/nature22321} {\bibfield  {journal} {\bibinfo
  {journal} {Nature}\ }\textbf {\bibinfo {volume} {545}},\ \bibinfo {pages}
  {327} (\bibinfo {year} {2017})}\BibitemShut {NoStop}%
\bibitem [{\citenamefont {Shi}\ and\ \citenamefont {Ma}(2013)}]{Shi}%
  \BibitemOpen
  \bibfield  {author} {\bibinfo {author} {\bibfnamefont {X.-q.}\ \bibnamefont
  {Shi}}\ and\ \bibinfo {author} {\bibfnamefont {Y.-q.}\ \bibnamefont {Ma}},\
  }\bibfield  {title} {\bibinfo {title} {Topological structure dynamics
  revealing collective evolution in active nematics},\ }\href@noop {}
  {\bibfield  {journal} {\bibinfo  {journal} {Nature Communications}\ }\textbf
  {\bibinfo {volume} {4}},\ \bibinfo {pages} {3013} (\bibinfo {year}
  {2013})}\BibitemShut {NoStop}%
\bibitem [{\citenamefont {Thampi}\ \emph {et~al.}(2014)\citenamefont {Thampi},
  \citenamefont {Golestanian},\ and\ \citenamefont
  {Yeomans}}]{ThampiTurbulence}%
  \BibitemOpen
  \bibfield  {author} {\bibinfo {author} {\bibfnamefont {S.~P.}\ \bibnamefont
  {Thampi}}, \bibinfo {author} {\bibfnamefont {R.}~\bibnamefont
  {Golestanian}},\ and\ \bibinfo {author} {\bibfnamefont {J.~M.}\ \bibnamefont
  {Yeomans}},\ }\bibfield  {title} {\bibinfo {title} {Instabilities and
  topological defects in active nematics},\ }\href
  {https://doi.org/10.1209/0295-5075/105/18001} {\bibfield  {journal} {\bibinfo
   {journal} {{EPL} (Europhysics Letters)}\ }\textbf {\bibinfo {volume}
  {105}},\ \bibinfo {pages} {18001} (\bibinfo {year} {2014})}\BibitemShut
  {NoStop}%
\bibitem [{\citenamefont {Giomi}\ \emph {et~al.}(2013)\citenamefont {Giomi},
  \citenamefont {Bowick}, \citenamefont {Ma},\ and\ \citenamefont
  {Marchetti}}]{DefectsGiomi}%
  \BibitemOpen
  \bibfield  {author} {\bibinfo {author} {\bibfnamefont {L.}~\bibnamefont
  {Giomi}}, \bibinfo {author} {\bibfnamefont {M.~J.}\ \bibnamefont {Bowick}},
  \bibinfo {author} {\bibfnamefont {X.}~\bibnamefont {Ma}},\ and\ \bibinfo
  {author} {\bibfnamefont {M.~C.}\ \bibnamefont {Marchetti}},\ }\bibfield
  {title} {\bibinfo {title} {Defect annihilation and proliferation in active
  nematics},\ }\href@noop {} {\bibfield  {journal} {\bibinfo  {journal} {Phys.
  Rev. Lett.}\ }\textbf {\bibinfo {volume} {110}},\ \bibinfo {pages} {228101}
  (\bibinfo {year} {2013})}\BibitemShut {NoStop}%
\bibitem [{\citenamefont {Shankar}\ \emph {et~al.}(2018)\citenamefont
  {Shankar}, \citenamefont {Ramaswamy}, \citenamefont {Marchetti},\ and\
  \citenamefont {Bowick}}]{SurajDefectUnbinding}%
  \BibitemOpen
  \bibfield  {author} {\bibinfo {author} {\bibfnamefont {S.}~\bibnamefont
  {Shankar}}, \bibinfo {author} {\bibfnamefont {S.}~\bibnamefont {Ramaswamy}},
  \bibinfo {author} {\bibfnamefont {M.~C.}\ \bibnamefont {Marchetti}},\ and\
  \bibinfo {author} {\bibfnamefont {M.~J.}\ \bibnamefont {Bowick}},\ }\bibfield
   {title} {\bibinfo {title} {Defect unbinding in active nematics},\ }\href
  {https://doi.org/10.1103/PhysRevLett.121.108002} {\bibfield  {journal}
  {\bibinfo  {journal} {Phys. Rev. Lett.}\ }\textbf {\bibinfo {volume} {121}},\
  \bibinfo {pages} {108002} (\bibinfo {year} {2018})}\BibitemShut {NoStop}%
\bibitem [{\citenamefont {Putzig}\ \emph {et~al.}(2016)\citenamefont {Putzig},
  \citenamefont {Redner}, \citenamefont {Baskaran},\ and\ \citenamefont
  {Baskaran}}]{Putzig2016SoftMatter}%
  \BibitemOpen
  \bibfield  {author} {\bibinfo {author} {\bibfnamefont {E.}~\bibnamefont
  {Putzig}}, \bibinfo {author} {\bibfnamefont {G.~S.}\ \bibnamefont {Redner}},
  \bibinfo {author} {\bibfnamefont {A.}~\bibnamefont {Baskaran}},\ and\
  \bibinfo {author} {\bibfnamefont {A.}~\bibnamefont {Baskaran}},\ }\bibfield
  {title} {\bibinfo {title} {Instabilities{,} defects{,} and defect ordering in
  an overdamped active nematic},\ }\href {https://doi.org/10.1039/C6SM00268D}
  {\bibfield  {journal} {\bibinfo  {journal} {Soft Matter}\ }\textbf {\bibinfo
  {volume} {12}},\ \bibinfo {pages} {3854} (\bibinfo {year}
  {2016})}\BibitemShut {NoStop}%
\bibitem [{\citenamefont {Oza}\ and\ \citenamefont {Dunkel}(2016)}]{Oza_2016}%
  \BibitemOpen
  \bibfield  {author} {\bibinfo {author} {\bibfnamefont {A.~U.}\ \bibnamefont
  {Oza}}\ and\ \bibinfo {author} {\bibfnamefont {J.}~\bibnamefont {Dunkel}},\
  }\bibfield  {title} {\bibinfo {title} {Antipolar ordering of topological
  defects in active liquid crystals},\ }\href
  {https://doi.org/10.1088/1367-2630/18/9/093006} {\bibfield  {journal}
  {\bibinfo  {journal} {New Journal of Physics}\ }\textbf {\bibinfo {volume}
  {18}},\ \bibinfo {pages} {093006} (\bibinfo {year} {2016})}\BibitemShut
  {NoStop}%
\bibitem [{\citenamefont {DeCamp}\ \emph {et~al.}(2015)\citenamefont {DeCamp},
  \citenamefont {Redner}, \citenamefont {Baskaran}, \citenamefont {Hagan},\
  and\ \citenamefont {Dogic}}]{DeCamp2015}%
  \BibitemOpen
  \bibfield  {author} {\bibinfo {author} {\bibfnamefont {S.~J.}\ \bibnamefont
  {DeCamp}}, \bibinfo {author} {\bibfnamefont {G.~S.}\ \bibnamefont {Redner}},
  \bibinfo {author} {\bibfnamefont {A.}~\bibnamefont {Baskaran}}, \bibinfo
  {author} {\bibfnamefont {M.~F.}\ \bibnamefont {Hagan}},\ and\ \bibinfo
  {author} {\bibfnamefont {Z.}~\bibnamefont {Dogic}},\ }\bibfield  {title}
  {\bibinfo {title} {Orientational order of motile defects in active
  nematics},\ }\href@noop {} {\bibfield  {journal} {\bibinfo  {journal} {Nature
  Materials}\ }\textbf {\bibinfo {volume} {14}},\ \bibinfo {pages} {1110 EP }
  (\bibinfo {year} {2015})}\BibitemShut {NoStop}%
\bibitem [{\citenamefont {Guillamat}\ \emph {et~al.}(2017)\citenamefont
  {Guillamat}, \citenamefont {Ign{\'e}s-Mullol},\ and\ \citenamefont
  {Sagu{\'e}s}}]{Pau2017}%
  \BibitemOpen
  \bibfield  {author} {\bibinfo {author} {\bibfnamefont {P.}~\bibnamefont
  {Guillamat}}, \bibinfo {author} {\bibfnamefont {J.}~\bibnamefont
  {Ign{\'e}s-Mullol}},\ and\ \bibinfo {author} {\bibfnamefont {F.}~\bibnamefont
  {Sagu{\'e}s}},\ }\bibfield  {title} {\bibinfo {title} {Taming active
  turbulence with patterned soft interfaces},\ }\href@noop {} {\bibfield
  {journal} {\bibinfo  {journal} {Nature Communications}\ }\textbf {\bibinfo
  {volume} {8}},\ \bibinfo {pages} {564} (\bibinfo {year} {2017})}\BibitemShut
  {NoStop}%
\bibitem [{\citenamefont {Kumar}\ \emph {et~al.}(2018)\citenamefont {Kumar},
  \citenamefont {Zhang}, \citenamefont {de~Pablo},\ and\ \citenamefont
  {Gardel}}]{GardelActinNematic}%
  \BibitemOpen
  \bibfield  {author} {\bibinfo {author} {\bibfnamefont {N.}~\bibnamefont
  {Kumar}}, \bibinfo {author} {\bibfnamefont {R.}~\bibnamefont {Zhang}},
  \bibinfo {author} {\bibfnamefont {J.~J.}\ \bibnamefont {de~Pablo}},\ and\
  \bibinfo {author} {\bibfnamefont {M.~L.}\ \bibnamefont {Gardel}},\ }\bibfield
   {title} {\bibinfo {title} {Tunable structure and dynamics of active liquid
  crystals},\ }\href@noop {} {\bibfield  {journal} {\bibinfo  {journal}
  {Science Advances}\ }\textbf {\bibinfo {volume} {4}} (\bibinfo {year}
  {2018})}\BibitemShut {NoStop}%
\bibitem [{\citenamefont {Lemma}\ \emph {et~al.}(2019)\citenamefont {Lemma},
  \citenamefont {DeCamp}, \citenamefont {You}, \citenamefont {Giomi},\ and\
  \citenamefont {Dogic}}]{VortexPaperLemma}%
  \BibitemOpen
  \bibfield  {author} {\bibinfo {author} {\bibfnamefont {L.~M.}\ \bibnamefont
  {Lemma}}, \bibinfo {author} {\bibfnamefont {S.~J.}\ \bibnamefont {DeCamp}},
  \bibinfo {author} {\bibfnamefont {Z.}~\bibnamefont {You}}, \bibinfo {author}
  {\bibfnamefont {L.}~\bibnamefont {Giomi}},\ and\ \bibinfo {author}
  {\bibfnamefont {Z.}~\bibnamefont {Dogic}},\ }\bibfield  {title} {\bibinfo
  {title} {Statistical properties of autonomous flows in 2d active nematics},\
  }\href@noop {} {\bibfield  {journal} {\bibinfo  {journal} {Soft Matter}\
  }\textbf {\bibinfo {volume} {15}},\ \bibinfo {pages} {3264} (\bibinfo {year}
  {2019})}\BibitemShut {NoStop}%
\bibitem [{\citenamefont {Tan}\ \emph {et~al.}(2019)\citenamefont {Tan},
  \citenamefont {Roberts}, \citenamefont {Smith}, \citenamefont {Olvera},
  \citenamefont {Arteaga}, \citenamefont {Fortini}, \citenamefont {Mitchell},\
  and\ \citenamefont {Hirst}}]{AmandaTan}%
  \BibitemOpen
  \bibfield  {author} {\bibinfo {author} {\bibfnamefont {A.~J.}\ \bibnamefont
  {Tan}}, \bibinfo {author} {\bibfnamefont {E.}~\bibnamefont {Roberts}},
  \bibinfo {author} {\bibfnamefont {S.~A.}\ \bibnamefont {Smith}}, \bibinfo
  {author} {\bibfnamefont {U.~A.}\ \bibnamefont {Olvera}}, \bibinfo {author}
  {\bibfnamefont {J.}~\bibnamefont {Arteaga}}, \bibinfo {author} {\bibfnamefont
  {S.}~\bibnamefont {Fortini}}, \bibinfo {author} {\bibfnamefont {K.~A.}\
  \bibnamefont {Mitchell}},\ and\ \bibinfo {author} {\bibfnamefont {L.~S.}\
  \bibnamefont {Hirst}},\ }\bibfield  {title} {\bibinfo {title} {{Topological
  chaos in active nematics}},\ }\href
  {https://doi.org/10.1038/s41567-019-0600-y} {\bibfield  {journal} {\bibinfo
  {journal} {Nature Physics}\ }\textbf {\bibinfo {volume} {15}},\ \bibinfo
  {pages} {1033} (\bibinfo {year} {2019})}\BibitemShut {NoStop}%
\bibitem [{\citenamefont {Shelley}(2016)}]{ShelleyReview}%
  \BibitemOpen
  \bibfield  {author} {\bibinfo {author} {\bibfnamefont {M.~J.}\ \bibnamefont
  {Shelley}},\ }\bibfield  {title} {\bibinfo {title} {The dynamics of
  microtubule/motor-protein assemblies in biology and physics},\ }\href
  {https://doi.org/10.1146/annurev-fluid-010814-013639} {\bibfield  {journal}
  {\bibinfo  {journal} {Annual Review of Fluid Mechanics}\ }\textbf {\bibinfo
  {volume} {48}},\ \bibinfo {pages} {487} (\bibinfo {year} {2016})},\ \Eprint
  {https://arxiv.org/abs/https://doi.org/10.1146/annurev-fluid-010814-013639}
  {https://doi.org/10.1146/annurev-fluid-010814-013639} \BibitemShut {NoStop}%
\bibitem [{\citenamefont {F{\"u}rthauer}\ \emph {et~al.}(2019)\citenamefont
  {F{\"u}rthauer}, \citenamefont {Lemma}, \citenamefont {Foster}, \citenamefont
  {Ems-McClung}, \citenamefont {Yu}, \citenamefont {Walczak}, \citenamefont
  {Dogic}, \citenamefont {Needleman},\ and\ \citenamefont
  {Shelley}}]{BezSebastian}%
  \BibitemOpen
  \bibfield  {author} {\bibinfo {author} {\bibfnamefont {S.}~\bibnamefont
  {F{\"u}rthauer}}, \bibinfo {author} {\bibfnamefont {B.}~\bibnamefont
  {Lemma}}, \bibinfo {author} {\bibfnamefont {P.~J.}\ \bibnamefont {Foster}},
  \bibinfo {author} {\bibfnamefont {S.~C.}\ \bibnamefont {Ems-McClung}},
  \bibinfo {author} {\bibfnamefont {C.-H.}\ \bibnamefont {Yu}}, \bibinfo
  {author} {\bibfnamefont {C.~E.}\ \bibnamefont {Walczak}}, \bibinfo {author}
  {\bibfnamefont {Z.}~\bibnamefont {Dogic}}, \bibinfo {author} {\bibfnamefont
  {D.~J.}\ \bibnamefont {Needleman}},\ and\ \bibinfo {author} {\bibfnamefont
  {M.~J.}\ \bibnamefont {Shelley}},\ }\bibfield  {title} {\bibinfo {title}
  {Self-straining of actively crosslinked microtubule networks},\ }\href@noop
  {} {\bibfield  {journal} {\bibinfo  {journal} {Nature Physics}\ } (\bibinfo
  {year} {2019})}\BibitemShut {NoStop}%
\bibitem [{\citenamefont {Kruse}\ and\ \citenamefont
  {J\"ulicher}(2000)}]{KruseContracting2000}%
  \BibitemOpen
  \bibfield  {author} {\bibinfo {author} {\bibfnamefont {K.}~\bibnamefont
  {Kruse}}\ and\ \bibinfo {author} {\bibfnamefont {F.}~\bibnamefont
  {J\"ulicher}},\ }\bibfield  {title} {\bibinfo {title} {Actively contracting
  bundles of polar filaments},\ }\href
  {https://doi.org/10.1103/PhysRevLett.85.1778} {\bibfield  {journal} {\bibinfo
   {journal} {Phys. Rev. Lett.}\ }\textbf {\bibinfo {volume} {85}},\ \bibinfo
  {pages} {1778} (\bibinfo {year} {2000})}\BibitemShut {NoStop}%
\bibitem [{\citenamefont {N{\'e}d{\'e}lec}\ and\ \citenamefont
  {Surrey}(2001)}]{Nedelec2001}%
  \BibitemOpen
  \bibfield  {author} {\bibinfo {author} {\bibfnamefont {F.}~\bibnamefont
  {N{\'e}d{\'e}lec}}\ and\ \bibinfo {author} {\bibfnamefont {T.}~\bibnamefont
  {Surrey}},\ }\bibfield  {title} {\bibinfo {title} {Dynamics of microtubule
  aster formation by motor complexes},\ }\href
  {https://doi.org/https://doi.org/10.1016/S1296-2147(01)01227-6} {\bibfield
  {journal} {\bibinfo  {journal} {Comptes Rendus de l'Acad{\'e}mie des
  Sciences}\ }\textbf {\bibinfo {volume} {2}},\ \bibinfo {pages} {841}
  (\bibinfo {year} {2001})}\BibitemShut {NoStop}%
\bibitem [{\citenamefont {Lenz}\ \emph {et~al.}(2012)\citenamefont {Lenz},
  \citenamefont {Thoresen}, \citenamefont {Gardel},\ and\ \citenamefont
  {Dinner}}]{GardelLenzContractile}%
  \BibitemOpen
  \bibfield  {author} {\bibinfo {author} {\bibfnamefont {M.}~\bibnamefont
  {Lenz}}, \bibinfo {author} {\bibfnamefont {T.}~\bibnamefont {Thoresen}},
  \bibinfo {author} {\bibfnamefont {M.~L.}\ \bibnamefont {Gardel}},\ and\
  \bibinfo {author} {\bibfnamefont {A.~R.}\ \bibnamefont {Dinner}},\ }\bibfield
   {title} {\bibinfo {title} {Contractile units in disordered actomyosin
  bundles arise from f-actin buckling},\ }\href
  {https://doi.org/10.1103/PhysRevLett.108.238107} {\bibfield  {journal}
  {\bibinfo  {journal} {Physical review letters}\ }\textbf {\bibinfo {volume}
  {108}},\ \bibinfo {pages} {238107} (\bibinfo {year} {2012})}\BibitemShut
  {NoStop}%
\bibitem [{\citenamefont {Needleman}\ and\ \citenamefont
  {Dogic}(2017)}]{NeedlemanDogicReview}%
  \BibitemOpen
  \bibfield  {author} {\bibinfo {author} {\bibfnamefont {D.}~\bibnamefont
  {Needleman}}\ and\ \bibinfo {author} {\bibfnamefont {Z.}~\bibnamefont
  {Dogic}},\ }\bibfield  {title} {\bibinfo {title} {Active matter at the
  interface between materials science and cell biology},\ }\href@noop {}
  {\bibfield  {journal} {\bibinfo  {journal} {Nature Reviews Materials}\
  }\textbf {\bibinfo {volume} {2}},\ \bibinfo {pages} {17048} (\bibinfo {year}
  {2017})}\BibitemShut {NoStop}%
\bibitem [{\citenamefont {Foster}\ \emph {et~al.}(2015)\citenamefont {Foster},
  \citenamefont {Furthauer}, \citenamefont {Shelley},\ and\ \citenamefont
  {Needleman}}]{Peter-eLife2015}%
  \BibitemOpen
  \bibfield  {author} {\bibinfo {author} {\bibfnamefont {P.~J.}\ \bibnamefont
  {Foster}}, \bibinfo {author} {\bibfnamefont {S.}~\bibnamefont {Furthauer}},
  \bibinfo {author} {\bibfnamefont {M.~J.}\ \bibnamefont {Shelley}},\ and\
  \bibinfo {author} {\bibfnamefont {D.~J.}\ \bibnamefont {Needleman}},\
  }\bibfield  {title} {\bibinfo {title} {Active contraction of microtubule
  networks},\ }\href@noop {} {\bibfield  {journal} {\bibinfo  {journal}
  {eLife}\ }\textbf {\bibinfo {volume} {4}} (\bibinfo {year}
  {2015})}\BibitemShut {NoStop}%
\bibitem [{\citenamefont {Howard}\ \emph {et~al.}(1989)\citenamefont {Howard},
  \citenamefont {Hudspeth},\ and\ \citenamefont {Vale}}]{HowardVale}%
  \BibitemOpen
  \bibfield  {author} {\bibinfo {author} {\bibfnamefont {J.}~\bibnamefont
  {Howard}}, \bibinfo {author} {\bibfnamefont {A.~J.}\ \bibnamefont
  {Hudspeth}},\ and\ \bibinfo {author} {\bibfnamefont {R.~D.}\ \bibnamefont
  {Vale}},\ }\bibfield  {title} {\bibinfo {title} {Movement of microtubules by
  single kinesin molecules},\ }\href {https://doi.org/10.1038/342154a0}
  {\bibfield  {journal} {\bibinfo  {journal} {Nature}\ }\textbf {\bibinfo
  {volume} {342}},\ \bibinfo {pages} {154} (\bibinfo {year}
  {1989})}\BibitemShut {NoStop}%
\bibitem [{\citenamefont {Svoboda}\ \emph {et~al.}(1993)\citenamefont
  {Svoboda}, \citenamefont {Schmidt}, \citenamefont {Schnapp},\ and\
  \citenamefont {Block}}]{StochasticKinesin}%
  \BibitemOpen
  \bibfield  {author} {\bibinfo {author} {\bibfnamefont {K.}~\bibnamefont
  {Svoboda}}, \bibinfo {author} {\bibfnamefont {C.~F.}\ \bibnamefont
  {Schmidt}}, \bibinfo {author} {\bibfnamefont {B.~J.}\ \bibnamefont
  {Schnapp}},\ and\ \bibinfo {author} {\bibfnamefont {S.~M.}\ \bibnamefont
  {Block}},\ }\bibfield  {title} {\bibinfo {title} {Direct observation of
  kinesin stepping by optical trapping interferometry},\ }\href
  {https://doi.org/10.1038/365721a0} {\bibfield  {journal} {\bibinfo  {journal}
  {Nature}\ }\textbf {\bibinfo {volume} {365}},\ \bibinfo {pages} {721}
  (\bibinfo {year} {1993})}\BibitemShut {NoStop}%
\bibitem [{\citenamefont {Visscher}\ \emph {et~al.}(1999)\citenamefont
  {Visscher}, \citenamefont {Schnitzer},\ and\ \citenamefont
  {Block}}]{SchnitzerForce}%
  \BibitemOpen
  \bibfield  {author} {\bibinfo {author} {\bibfnamefont {K.}~\bibnamefont
  {Visscher}}, \bibinfo {author} {\bibfnamefont {M.~J.}\ \bibnamefont
  {Schnitzer}},\ and\ \bibinfo {author} {\bibfnamefont {S.~M.}\ \bibnamefont
  {Block}},\ }\bibfield  {title} {\bibinfo {title} {Single kinesin molecules
  studied with a molecular force clamp},\ }\href@noop {} {\bibfield  {journal}
  {\bibinfo  {journal} {Nature}\ }\textbf {\bibinfo {volume} {400}},\ \bibinfo
  {pages} {184} (\bibinfo {year} {1999})}\BibitemShut {NoStop}%
\bibitem [{\citenamefont {Bieling}\ \emph {et~al.}(2008)\citenamefont
  {Bieling}, \citenamefont {Telley}, \citenamefont {Piehler},\ and\
  \citenamefont {Surrey}}]{KinesinMultiMotor}%
  \BibitemOpen
  \bibfield  {author} {\bibinfo {author} {\bibfnamefont {P.}~\bibnamefont
  {Bieling}}, \bibinfo {author} {\bibfnamefont {I.~A.}\ \bibnamefont {Telley}},
  \bibinfo {author} {\bibfnamefont {J.}~\bibnamefont {Piehler}},\ and\ \bibinfo
  {author} {\bibfnamefont {T.}~\bibnamefont {Surrey}},\ }\bibfield  {title}
  {\bibinfo {title} {Processive kinesins require loose mechanical coupling for
  efficient collective motility},\ }\href
  {https://doi.org/10.1038/embor.2008.169} {\bibfield  {journal} {\bibinfo
  {journal} {EMBO reports}\ }\textbf {\bibinfo {volume} {9}},\ \bibinfo {pages}
  {1121} (\bibinfo {year} {2008})}\BibitemShut {NoStop}%
\bibitem [{\citenamefont {Khataee}\ and\ \citenamefont
  {Howard}(2019)}]{HowardPRL}%
  \BibitemOpen
  \bibfield  {author} {\bibinfo {author} {\bibfnamefont {H.}~\bibnamefont
  {Khataee}}\ and\ \bibinfo {author} {\bibfnamefont {J.}~\bibnamefont
  {Howard}},\ }\bibfield  {title} {\bibinfo {title} {Force generated by two
  kinesin motors depends on the load direction and intermolecular coupling},\
  }\href {https://doi.org/10.1103/PhysRevLett.122.188101} {\bibfield  {journal}
  {\bibinfo  {journal} {Phys. Rev. Lett.}\ }\textbf {\bibinfo {volume} {122}},\
  \bibinfo {pages} {188101} (\bibinfo {year} {2019})}\BibitemShut {NoStop}%
\bibitem [{\citenamefont {Kruse}\ \emph {et~al.}(2004)\citenamefont {Kruse},
  \citenamefont {Joanny}, \citenamefont {J\"ulicher}, \citenamefont {Prost},\
  and\ \citenamefont {Sekimoto}}]{KrusePolar}%
  \BibitemOpen
  \bibfield  {author} {\bibinfo {author} {\bibfnamefont {K.}~\bibnamefont
  {Kruse}}, \bibinfo {author} {\bibfnamefont {J.~F.}\ \bibnamefont {Joanny}},
  \bibinfo {author} {\bibfnamefont {F.}~\bibnamefont {J\"ulicher}}, \bibinfo
  {author} {\bibfnamefont {J.}~\bibnamefont {Prost}},\ and\ \bibinfo {author}
  {\bibfnamefont {K.}~\bibnamefont {Sekimoto}},\ }\bibfield  {title} {\bibinfo
  {title} {Asters, vortices, and rotating spirals in active gels of polar
  filaments},\ }\href {https://doi.org/10.1103/PhysRevLett.92.078101}
  {\bibfield  {journal} {\bibinfo  {journal} {Phys. Rev. Lett.}\ }\textbf
  {\bibinfo {volume} {92}},\ \bibinfo {pages} {078101} (\bibinfo {year}
  {2004})}\BibitemShut {NoStop}%
\bibitem [{\citenamefont {Gao}\ \emph {et~al.}(2015)\citenamefont {Gao},
  \citenamefont {Blackwell}, \citenamefont {Glaser}, \citenamefont
  {Betterton},\ and\ \citenamefont {Shelley}}]{GaoBetterton}%
  \BibitemOpen
  \bibfield  {author} {\bibinfo {author} {\bibfnamefont {T.}~\bibnamefont
  {Gao}}, \bibinfo {author} {\bibfnamefont {R.}~\bibnamefont {Blackwell}},
  \bibinfo {author} {\bibfnamefont {M.~A.}\ \bibnamefont {Glaser}}, \bibinfo
  {author} {\bibfnamefont {M.~D.}\ \bibnamefont {Betterton}},\ and\ \bibinfo
  {author} {\bibfnamefont {M.~J.}\ \bibnamefont {Shelley}},\ }\bibfield
  {title} {\bibinfo {title} {Multiscale polar theory of microtubule and
  motor-protein assemblies},\ }\href@noop {} {\bibfield  {journal} {\bibinfo
  {journal} {Phys. Rev. Lett.}\ }\textbf {\bibinfo {volume} {114}},\ \bibinfo
  {pages} {048101} (\bibinfo {year} {2015})}\BibitemShut {NoStop}%
\bibitem [{\citenamefont {Blackwell}\ \emph {et~al.}(2016)\citenamefont
  {Blackwell}, \citenamefont {Sweezy-Schindler}, \citenamefont {Baldwin},
  \citenamefont {Hough}, \citenamefont {Glaser},\ and\ \citenamefont
  {Betterton}}]{BettertonSoftMatter}%
  \BibitemOpen
  \bibfield  {author} {\bibinfo {author} {\bibfnamefont {R.}~\bibnamefont
  {Blackwell}}, \bibinfo {author} {\bibfnamefont {O.}~\bibnamefont
  {Sweezy-Schindler}}, \bibinfo {author} {\bibfnamefont {C.}~\bibnamefont
  {Baldwin}}, \bibinfo {author} {\bibfnamefont {L.~E.}\ \bibnamefont {Hough}},
  \bibinfo {author} {\bibfnamefont {M.~A.}\ \bibnamefont {Glaser}},\ and\
  \bibinfo {author} {\bibfnamefont {M.~D.}\ \bibnamefont {Betterton}},\
  }\bibfield  {title} {\bibinfo {title} {Microscopic origins of anisotropic
  active stress in motor-driven nematic liquid crystals},\ }\href
  {https://doi.org/10.1039/C5SM02506K} {\bibfield  {journal} {\bibinfo
  {journal} {Soft Matter}\ }\textbf {\bibinfo {volume} {12}},\ \bibinfo {pages}
  {2676} (\bibinfo {year} {2016})}\BibitemShut {NoStop}%
\bibitem [{\citenamefont {Striebel}\ \emph {et~al.}(2020)\citenamefont
  {Striebel}, \citenamefont {Graf},\ and\ \citenamefont {Frey}}]{FreyModel}%
  \BibitemOpen
  \bibfield  {author} {\bibinfo {author} {\bibfnamefont {M.}~\bibnamefont
  {Striebel}}, \bibinfo {author} {\bibfnamefont {I.~R.}\ \bibnamefont {Graf}},\
  and\ \bibinfo {author} {\bibfnamefont {E.}~\bibnamefont {Frey}},\ }\bibfield
  {title} {\bibinfo {title} {A mechanistic view of collective filament motion
  in active nematic networks},\ }\href
  {https://doi.org/https://doi.org/10.1016/j.bpj.2019.11.3387} {\bibfield
  {journal} {\bibinfo  {journal} {Biophysical Journal}\ }\textbf {\bibinfo
  {volume} {118}},\ \bibinfo {pages} {313} (\bibinfo {year}
  {2020})}\BibitemShut {NoStop}%
\bibitem [{\citenamefont {Hilitski}\ \emph {et~al.}(2015)\citenamefont
  {Hilitski}, \citenamefont {Ward}, \citenamefont {Cajamarca}, \citenamefont
  {Hagan}, \citenamefont {Grason},\ and\ \citenamefont
  {Dogic}}]{FeodorFriction}%
  \BibitemOpen
  \bibfield  {author} {\bibinfo {author} {\bibfnamefont {F.}~\bibnamefont
  {Hilitski}}, \bibinfo {author} {\bibfnamefont {A.~R.}\ \bibnamefont {Ward}},
  \bibinfo {author} {\bibfnamefont {L.}~\bibnamefont {Cajamarca}}, \bibinfo
  {author} {\bibfnamefont {M.~F.}\ \bibnamefont {Hagan}}, \bibinfo {author}
  {\bibfnamefont {G.~M.}\ \bibnamefont {Grason}},\ and\ \bibinfo {author}
  {\bibfnamefont {Z.}~\bibnamefont {Dogic}},\ }\bibfield  {title} {\bibinfo
  {title} {Measuring cohesion between macromolecular filaments one pair at a
  time: Depletion-induced microtubule bundling},\ }\href@noop {} {\bibfield
  {journal} {\bibinfo  {journal} {Phys. Rev. Lett.}\ }\textbf {\bibinfo
  {volume} {114}},\ \bibinfo {pages} {138102} (\bibinfo {year}
  {2015})}\BibitemShut {NoStop}%
\bibitem [{\citenamefont {Sanchez}\ \emph {et~al.}(2012)\citenamefont
  {Sanchez}, \citenamefont {Chen}, \citenamefont {DeCamp}, \citenamefont
  {Heymann},\ and\ \citenamefont {Dogic}}]{Sanchez2012}%
  \BibitemOpen
  \bibfield  {author} {\bibinfo {author} {\bibfnamefont {T.}~\bibnamefont
  {Sanchez}}, \bibinfo {author} {\bibfnamefont {D.~T.~N.}\ \bibnamefont
  {Chen}}, \bibinfo {author} {\bibfnamefont {S.~J.}\ \bibnamefont {DeCamp}},
  \bibinfo {author} {\bibfnamefont {M.}~\bibnamefont {Heymann}},\ and\ \bibinfo
  {author} {\bibfnamefont {Z.}~\bibnamefont {Dogic}},\ }\bibfield  {title}
  {\bibinfo {title} {Spontaneous motion in hierarchically assembled active
  matter},\ }\href@noop {} {\bibfield  {journal} {\bibinfo  {journal} {Nature}\
  }\textbf {\bibinfo {volume} {491}},\ \bibinfo {pages} {431 EP } (\bibinfo
  {year} {2012})}\BibitemShut {NoStop}%
\bibitem [{\citenamefont {Henkin}\ \emph {et~al.}(2014)\citenamefont {Henkin},
  \citenamefont {DeCamp}, \citenamefont {Chen}, \citenamefont {Sanchez},\ and\
  \citenamefont {Dogic}}]{Henkin2014}%
  \BibitemOpen
  \bibfield  {author} {\bibinfo {author} {\bibfnamefont {G.}~\bibnamefont
  {Henkin}}, \bibinfo {author} {\bibfnamefont {S.~J.}\ \bibnamefont {DeCamp}},
  \bibinfo {author} {\bibfnamefont {D.~T.~N.}\ \bibnamefont {Chen}}, \bibinfo
  {author} {\bibfnamefont {T.}~\bibnamefont {Sanchez}},\ and\ \bibinfo {author}
  {\bibfnamefont {Z.}~\bibnamefont {Dogic}},\ }\bibfield  {title} {\bibinfo
  {title} {Tunable dynamics of microtubule-based active isotropic gels},\
  }\href@noop {} {\bibfield  {journal} {\bibinfo  {journal} {Philosophical
  Transactions of the Royal Society A: Mathematical, Physical and Engineering
  Sciences}\ }\textbf {\bibinfo {volume} {372}},\ \bibinfo {pages} {20140142}
  (\bibinfo {year} {2014})}\BibitemShut {NoStop}%
\bibitem [{\citenamefont {Chandrakar}\ \emph {et~al.}(2018)\citenamefont
  {Chandrakar}, \citenamefont {Berezney}, \citenamefont {Lemma}, \citenamefont
  {Hishamunda}, \citenamefont {Berry}, \citenamefont {Wu}, \citenamefont
  {Subramanian}, \citenamefont {Chung}, \citenamefont {Needleman},
  \citenamefont {Gelles},\ and\ \citenamefont {Dogic}}]{Pooja-arxiv}%
  \BibitemOpen
  \bibfield  {author} {\bibinfo {author} {\bibfnamefont {P.}~\bibnamefont
  {Chandrakar}}, \bibinfo {author} {\bibfnamefont {J.}~\bibnamefont
  {Berezney}}, \bibinfo {author} {\bibfnamefont {B.}~\bibnamefont {Lemma}},
  \bibinfo {author} {\bibfnamefont {B.}~\bibnamefont {Hishamunda}}, \bibinfo
  {author} {\bibfnamefont {A.}~\bibnamefont {Berry}}, \bibinfo {author}
  {\bibfnamefont {K.-T.}\ \bibnamefont {Wu}}, \bibinfo {author} {\bibfnamefont
  {R.}~\bibnamefont {Subramanian}}, \bibinfo {author} {\bibfnamefont
  {J.}~\bibnamefont {Chung}}, \bibinfo {author} {\bibfnamefont
  {D.}~\bibnamefont {Needleman}}, \bibinfo {author} {\bibfnamefont
  {J.}~\bibnamefont {Gelles}},\ and\ \bibinfo {author} {\bibfnamefont
  {Z.}~\bibnamefont {Dogic}},\ }\href@noop {} {\bibinfo {title}
  {Microtubule-based active fluids with improved lifetime, temporal stability
  and miscibility with passive soft materials}} (\bibinfo {year} {2018}),\
  \Eprint {https://arxiv.org/abs/1811.05026} {arXiv:1811.05026 [cond-mat.soft]}
  \BibitemShut {NoStop}%
\bibitem [{\citenamefont {Duclos}\ \emph {et~al.}(2020)\citenamefont {Duclos},
  \citenamefont {Adkins}, \citenamefont {Banerjee}, \citenamefont {Peterson},
  \citenamefont {Varghese}, \citenamefont {Kolvin}, \citenamefont {Baskaran},
  \citenamefont {Pelcovits}, \citenamefont {Powers}, \citenamefont {Baskaran},
  \citenamefont {Toschi}, \citenamefont {Hagan}, \citenamefont {Streichan},
  \citenamefont {Vitelli}, \citenamefont {Beller},\ and\ \citenamefont
  {Dogic}}]{DuclosAdkins3D}%
  \BibitemOpen
  \bibfield  {author} {\bibinfo {author} {\bibfnamefont {G.}~\bibnamefont
  {Duclos}}, \bibinfo {author} {\bibfnamefont {R.}~\bibnamefont {Adkins}},
  \bibinfo {author} {\bibfnamefont {D.}~\bibnamefont {Banerjee}}, \bibinfo
  {author} {\bibfnamefont {M.~S.~E.}\ \bibnamefont {Peterson}}, \bibinfo
  {author} {\bibfnamefont {M.}~\bibnamefont {Varghese}}, \bibinfo {author}
  {\bibfnamefont {I.}~\bibnamefont {Kolvin}}, \bibinfo {author} {\bibfnamefont
  {A.}~\bibnamefont {Baskaran}}, \bibinfo {author} {\bibfnamefont {R.~A.}\
  \bibnamefont {Pelcovits}}, \bibinfo {author} {\bibfnamefont {T.~R.}\
  \bibnamefont {Powers}}, \bibinfo {author} {\bibfnamefont {A.}~\bibnamefont
  {Baskaran}}, \bibinfo {author} {\bibfnamefont {F.}~\bibnamefont {Toschi}},
  \bibinfo {author} {\bibfnamefont {M.~F.}\ \bibnamefont {Hagan}}, \bibinfo
  {author} {\bibfnamefont {S.~J.}\ \bibnamefont {Streichan}}, \bibinfo {author}
  {\bibfnamefont {V.}~\bibnamefont {Vitelli}}, \bibinfo {author} {\bibfnamefont
  {D.~A.}\ \bibnamefont {Beller}},\ and\ \bibinfo {author} {\bibfnamefont
  {Z.}~\bibnamefont {Dogic}},\ }\bibfield  {title} {\bibinfo {title}
  {Topological structure and dynamics of three-dimensional active nematics},\
  }\href {https://doi.org/10.1126/science.aaz4547} {\bibfield  {journal}
  {\bibinfo  {journal} {Science}\ }\textbf {\bibinfo {volume} {367}},\ \bibinfo
  {pages} {1120} (\bibinfo {year} {2020})}
  \BibitemShut
  {NoStop}%
\bibitem [{\citenamefont {Berliner}\ \emph {et~al.}(1995)\citenamefont
  {Berliner}, \citenamefont {Young}, \citenamefont {Anderson}, \citenamefont
  {Mahtani},\ and\ \citenamefont {Gelles}}]{Berliner}%
  \BibitemOpen
  \bibfield  {author} {\bibinfo {author} {\bibfnamefont {E.}~\bibnamefont
  {Berliner}}, \bibinfo {author} {\bibfnamefont {E.~C.}\ \bibnamefont {Young}},
  \bibinfo {author} {\bibfnamefont {K.}~\bibnamefont {Anderson}}, \bibinfo
  {author} {\bibfnamefont {H.~K.}\ \bibnamefont {Mahtani}},\ and\ \bibinfo
  {author} {\bibfnamefont {J.}~\bibnamefont {Gelles}},\ }\bibfield  {title}
  {\bibinfo {title} {Failure of a single-headed kinesin to track parallel to
  microtubule protofilaments},\ }\href {https://doi.org/10.1038/373718a0}
  {\bibfield  {journal} {\bibinfo  {journal} {Nature}\ }\textbf {\bibinfo
  {volume} {373}},\ \bibinfo {pages} {718} (\bibinfo {year}
  {1995})}\BibitemShut {NoStop}%
\bibitem [{\citenamefont {Ward}\ \emph {et~al.}(2015)\citenamefont {Ward},
  \citenamefont {Hilitski}, \citenamefont {Schwenger}, \citenamefont {Welch},
  \citenamefont {Lau}, \citenamefont {Vitelli}, \citenamefont {Mahadevan},\
  and\ \citenamefont {Dogic}}]{Ward2015}%
  \BibitemOpen
  \bibfield  {author} {\bibinfo {author} {\bibfnamefont {A.}~\bibnamefont
  {Ward}}, \bibinfo {author} {\bibfnamefont {F.}~\bibnamefont {Hilitski}},
  \bibinfo {author} {\bibfnamefont {W.}~\bibnamefont {Schwenger}}, \bibinfo
  {author} {\bibfnamefont {D.}~\bibnamefont {Welch}}, \bibinfo {author}
  {\bibfnamefont {A.~W.~C.}\ \bibnamefont {Lau}}, \bibinfo {author}
  {\bibfnamefont {V.}~\bibnamefont {Vitelli}}, \bibinfo {author} {\bibfnamefont
  {L.}~\bibnamefont {Mahadevan}},\ and\ \bibinfo {author} {\bibfnamefont
  {Z.}~\bibnamefont {Dogic}},\ }\bibfield  {title} {\bibinfo {title} {Solid
  friction between soft filaments},\ }\href@noop {} {\bibfield  {journal}
  {\bibinfo  {journal} {Nature Materials}\ }\textbf {\bibinfo {volume} {14}},\
  \bibinfo {pages} {583 EP } (\bibinfo {year} {2015})}\BibitemShut {NoStop}%
\bibitem [{\citenamefont {Dell'Arciprete}\ \emph {et~al.}(2018)\citenamefont
  {Dell'Arciprete}, \citenamefont {Blow}, \citenamefont {Brown}, \citenamefont
  {Farrell}, \citenamefont {Lintuvuori}, \citenamefont {McVey}, \citenamefont
  {Marenduzzo},\ and\ \citenamefont {Poon}}]{Poon}%
  \BibitemOpen
  \bibfield  {author} {\bibinfo {author} {\bibfnamefont {D.}~\bibnamefont
  {Dell'Arciprete}}, \bibinfo {author} {\bibfnamefont {M.~L.}\ \bibnamefont
  {Blow}}, \bibinfo {author} {\bibfnamefont {A.~T.}\ \bibnamefont {Brown}},
  \bibinfo {author} {\bibfnamefont {F.~D.~C.}\ \bibnamefont {Farrell}},
  \bibinfo {author} {\bibfnamefont {J.~S.}\ \bibnamefont {Lintuvuori}},
  \bibinfo {author} {\bibfnamefont {A.~F.}\ \bibnamefont {McVey}}, \bibinfo
  {author} {\bibfnamefont {D.}~\bibnamefont {Marenduzzo}},\ and\ \bibinfo
  {author} {\bibfnamefont {W.~C.~K.}\ \bibnamefont {Poon}},\ }\bibfield
  {title} {\bibinfo {title} {A growing bacterial colony in two dimensions as an
  active nematic},\ }\href {https://doi.org/10.1038/s41467-018-06370-3}
  {\bibfield  {journal} {\bibinfo  {journal} {Nature Communications}\ }\textbf
  {\bibinfo {volume} {9}},\ \bibinfo {pages} {4190} (\bibinfo {year}
  {2018})}\BibitemShut {NoStop}%
\bibitem [{\citenamefont {Qian}\ \emph {et~al.}(1991)\citenamefont {Qian},
  \citenamefont {Sheetz},\ and\ \citenamefont {Elson}}]{Qian:1991}%
  \BibitemOpen
  \bibfield  {author} {\bibinfo {author} {\bibfnamefont {H.}~\bibnamefont
  {Qian}}, \bibinfo {author} {\bibfnamefont {M.~P.}\ \bibnamefont {Sheetz}},\
  and\ \bibinfo {author} {\bibfnamefont {E.~L.}\ \bibnamefont {Elson}},\
  }\bibfield  {title} {\bibinfo {title} {Single particle tracking. analysis of
  diffusion and flow in two-dimensional systems},\ }\href
  {https://doi.org/10.1016/S0006-3495(91)82125-7} {\bibfield  {journal}
  {\bibinfo  {journal} {Biophysical journal}\ }\textbf {\bibinfo {volume}
  {60}},\ \bibinfo {pages} {910} (\bibinfo {year} {1991})}\BibitemShut
  {NoStop}%
\bibitem [{\citenamefont {Giomi}(2015)}]{LucaGeometry}%
  \BibitemOpen
  \bibfield  {author} {\bibinfo {author} {\bibfnamefont {L.}~\bibnamefont
  {Giomi}},\ }\bibfield  {title} {\bibinfo {title} {Geometry and topology of
  turbulence in active nematics},\ }\href@noop {} {\bibfield  {journal}
  {\bibinfo  {journal} {Phys. Rev. X}\ }\textbf {\bibinfo {volume} {5}},\
  \bibinfo {pages} {031003} (\bibinfo {year} {2015})}\BibitemShut {NoStop}%
\bibitem [{\citenamefont {Mart{\'\i}nez-Prat}\ \emph
  {et~al.}(2021)\citenamefont {Mart{\'\i}nez-Prat}, \citenamefont {Alert},
  \citenamefont {Meng}, \citenamefont {Ign{\'e}s-Mullol}, \citenamefont
  {Joanny}, \citenamefont {Casademunt}, \citenamefont {Golestanian},\ and\
  \citenamefont {Sagu{\'e}s}}]{RAlertActiveTurb}%
  \BibitemOpen
  \bibfield  {author} {\bibinfo {author} {\bibfnamefont {B.}~\bibnamefont
  {Mart{\'\i}nez-Prat}}, \bibinfo {author} {\bibfnamefont {R.}~\bibnamefont
  {Alert}}, \bibinfo {author} {\bibfnamefont {F.}~\bibnamefont {Meng}},
  \bibinfo {author} {\bibfnamefont {J.}~\bibnamefont {Ign{\'e}s-Mullol}},
  \bibinfo {author} {\bibfnamefont {J.-F.}\ \bibnamefont {Joanny}}, \bibinfo
  {author} {\bibfnamefont {J.}~\bibnamefont {Casademunt}}, \bibinfo {author}
  {\bibfnamefont {R.}~\bibnamefont {Golestanian}},\ and\ \bibinfo {author}
  {\bibfnamefont {F.}~\bibnamefont {Sagu{\'e}s}},\ }\href@noop {} {\bibinfo
  {title} {Scaling regimes of active turbulence with external dissipation}}
  (\bibinfo {year} {2021}),\ \Eprint {https://arxiv.org/abs/2101.11570}
  {arXiv:2101.11570} \BibitemShut {NoStop}%
\bibitem [{\citenamefont {Norton}\ \emph {et~al.}(2018)\citenamefont {Norton},
  \citenamefont {Baskaran}, \citenamefont {Opathalage}, \citenamefont
  {Langeslay}, \citenamefont {Fraden}, \citenamefont {Baskaran},\ and\
  \citenamefont {Hagan}}]{NortonTheory}%
  \BibitemOpen
  \bibfield  {author} {\bibinfo {author} {\bibfnamefont {M.~M.}\ \bibnamefont
  {Norton}}, \bibinfo {author} {\bibfnamefont {A.}~\bibnamefont {Baskaran}},
  \bibinfo {author} {\bibfnamefont {A.}~\bibnamefont {Opathalage}}, \bibinfo
  {author} {\bibfnamefont {B.}~\bibnamefont {Langeslay}}, \bibinfo {author}
  {\bibfnamefont {S.}~\bibnamefont {Fraden}}, \bibinfo {author} {\bibfnamefont
  {A.}~\bibnamefont {Baskaran}},\ and\ \bibinfo {author} {\bibfnamefont
  {M.~F.}\ \bibnamefont {Hagan}},\ }\bibfield  {title} {\bibinfo {title}
  {Insensitivity of active nematic liquid crystal dynamics to topological
  constraints},\ }\href {https://doi.org/10.1103/PhysRevE.97.012702} {\bibfield
   {journal} {\bibinfo  {journal} {Phys. Rev. E}\ }\textbf {\bibinfo {volume}
  {97}},\ \bibinfo {pages} {012702} (\bibinfo {year} {2018})}\BibitemShut
  {NoStop}%
\bibitem [{\citenamefont {Opathalage}\ \emph {et~al.}(2019)\citenamefont
  {Opathalage}, \citenamefont {Norton}, \citenamefont {Juniper}, \citenamefont
  {Langeslay}, \citenamefont {Aghvami}, \citenamefont {Fraden},\ and\
  \citenamefont {Dogic}}]{AchiniMike}%
  \BibitemOpen
  \bibfield  {author} {\bibinfo {author} {\bibfnamefont {A.}~\bibnamefont
  {Opathalage}}, \bibinfo {author} {\bibfnamefont {M.~M.}\ \bibnamefont
  {Norton}}, \bibinfo {author} {\bibfnamefont {M.~P.~N.}\ \bibnamefont
  {Juniper}}, \bibinfo {author} {\bibfnamefont {B.}~\bibnamefont {Langeslay}},
  \bibinfo {author} {\bibfnamefont {S.~A.}\ \bibnamefont {Aghvami}}, \bibinfo
  {author} {\bibfnamefont {S.}~\bibnamefont {Fraden}},\ and\ \bibinfo {author}
  {\bibfnamefont {Z.}~\bibnamefont {Dogic}},\ }\bibfield  {title} {\bibinfo
  {title} {Self-organized dynamics and the transition to turbulence of confined
  active nematics},\ }\href@noop {} {\bibfield  {journal} {\bibinfo  {journal}
  {Proceedings of the National Academy of Sciences}\ }\textbf {\bibinfo
  {volume} {116}},\ \bibinfo {pages} {4788} (\bibinfo {year}
  {2019})}\BibitemShut {NoStop}%
\bibitem [{\citenamefont {Tayar}\ \emph {et~al.}(2021)\citenamefont {Tayar},
  \citenamefont {Hagan},\ and\ \citenamefont {Dogic}}]{AlexPNAS2021}%
  \BibitemOpen
  \bibfield  {author} {\bibinfo {author} {\bibfnamefont {A.~M.}\ \bibnamefont
  {Tayar}}, \bibinfo {author} {\bibfnamefont {M.~F.}\ \bibnamefont {Hagan}},\
  and\ \bibinfo {author} {\bibfnamefont {Z.}~\bibnamefont {Dogic}},\ }\bibfield
   {title} {\bibinfo {title} {Active liquid crystals powered by force-sensing
  dna-motor clusters},\ }\bibfield  {journal} {\bibinfo  {journal} {Proceedings
  of the National Academy of Sciences}\ }\textbf {\bibinfo {volume} {118}},\
  \href {https://doi.org/10.1073/pnas.2102873118} {10.1073/pnas.2102873118}
  (\bibinfo {year} {2021}) \BibitemShut
  {NoStop}%
\bibitem [{\citenamefont {Liverpool}\ and\ \citenamefont
  {Marchetti}(2005)}]{Liverpool_2005}%
  \BibitemOpen
  \bibfield  {author} {\bibinfo {author} {\bibfnamefont {T.~B.}\ \bibnamefont
  {Liverpool}}\ and\ \bibinfo {author} {\bibfnamefont {M.~C.}\ \bibnamefont
  {Marchetti}},\ }\bibfield  {title} {\bibinfo {title} {Bridging the
  microscopic and the hydrodynamic in active filament solutions},\ }\href
  {https://doi.org/10.1209/epl/i2004-10414-0} {\bibfield  {journal} {\bibinfo
  {journal} {Europhysics Letters (EPL)}\ }\textbf {\bibinfo {volume} {69}},\
  \bibinfo {pages} {846–852} (\bibinfo {year} {2005})}\BibitemShut {NoStop}%
\bibitem [{\citenamefont {Belmonte}\ \emph {et~al.}(2017)\citenamefont
  {Belmonte}, \citenamefont {Leptin},\ and\ \citenamefont
  {N{\'e}d{\'e}lec}}]{Belmonte2017}%
  \BibitemOpen
  \bibfield  {author} {\bibinfo {author} {\bibfnamefont {J.~M.}\ \bibnamefont
  {Belmonte}}, \bibinfo {author} {\bibfnamefont {M.}~\bibnamefont {Leptin}},\
  and\ \bibinfo {author} {\bibfnamefont {F.}~\bibnamefont {N{\'e}d{\'e}lec}},\
  }\bibfield  {title} {\bibinfo {title} {A theory that predicts behaviors of
  disordered cytoskeletal networks},\ }\href
  {https://doi.org/10.15252/msb.20177796} {\bibfield  {journal} {\bibinfo
  {journal} {Molecular systems biology}\ }\textbf {\bibinfo {volume} {13}},\
  \bibinfo {pages} {941} (\bibinfo {year} {2017})}\BibitemShut {NoStop}%
\bibitem [{\citenamefont {Lenz}(2020)}]{LenzeLife2020}%
  \BibitemOpen
  \bibfield  {author} {\bibinfo {author} {\bibfnamefont {M.}~\bibnamefont
  {Lenz}},\ }\bibfield  {title} {\bibinfo {title} {Reversal of contractility as
  a signature of self-organization in cytoskeletal bundles},\ }\href
  {https://doi.org/10.7554/eLife.51751} {\bibfield  {journal} {\bibinfo
  {journal} {eLife}\ }\textbf {\bibinfo {volume} {9}},\ \bibinfo {pages}
  {e51751} (\bibinfo {year} {2020})}\BibitemShut {NoStop}%
\bibitem [{\citenamefont {Yu}\ \emph {et~al.}(2014)\citenamefont {Yu},
  \citenamefont {Langowitz}, \citenamefont {Wu}, \citenamefont {Brugues},\ and\
  \citenamefont {Needleman}}]{Che-HangSHG}%
  \BibitemOpen
  \bibfield  {author} {\bibinfo {author} {\bibfnamefont {C.-H.}\ \bibnamefont
  {Yu}}, \bibinfo {author} {\bibfnamefont {N.}~\bibnamefont {Langowitz}},
  \bibinfo {author} {\bibfnamefont {H.-Y.}\ \bibnamefont {Wu}}, \bibinfo
  {author} {\bibfnamefont {J.}~\bibnamefont {Brugues}},\ and\ \bibinfo {author}
  {\bibfnamefont {D.}~\bibnamefont {Needleman}},\ }\bibfield  {title} {\bibinfo
  {title} {Measuring microtubule polarity in spindles with
  second-harmonic-generation microscopy},\ }\href@noop {} {\bibfield  {journal}
  {\bibinfo  {journal} {Biophysical Journal}\ }\textbf {\bibinfo {volume}
  {106}},\ \bibinfo {pages} {1578} (\bibinfo {year} {2014})}\BibitemShut
  {NoStop}%
\bibitem [{SI()}]{SI}%
  \BibitemOpen
  \href@noop {} {\bibinfo {title} {See supplementary materials at [url] for
  detailed experimental and analysis methods, theoretical description; includes
  refs. [52-56].}}\BibitemShut {Stop}%
\bibitem [{\citenamefont {Subramanian}\ and\ \citenamefont
  {Gelles}(2007)}]{Subramanian2007}%
  \BibitemOpen
  \bibfield  {author} {\bibinfo {author} {\bibfnamefont {R.}~\bibnamefont
  {Subramanian}}\ and\ \bibinfo {author} {\bibfnamefont {J.}~\bibnamefont
  {Gelles}},\ }\bibfield  {title} {\bibinfo {title} {Two distinct modes of
  processive kinesin movement in mixtures of atp and amp-pnp},\ }\href@noop {}
  {\bibfield  {journal} {\bibinfo  {journal} {The Journal of General
  Physiology}\ }\textbf {\bibinfo {volume} {130}},\ \bibinfo {pages} {445}
  (\bibinfo {year} {2007})}\BibitemShut {NoStop}%
\bibitem [{\citenamefont {Castoldi}\ and\ \citenamefont
  {Popov}(2003)}]{PopovTubulin}%
  \BibitemOpen
  \bibfield  {author} {\bibinfo {author} {\bibfnamefont {M.}~\bibnamefont
  {Castoldi}}\ and\ \bibinfo {author} {\bibfnamefont {A.~V.}\ \bibnamefont
  {Popov}},\ }\bibfield  {title} {\bibinfo {title} {Purification of brain
  tubulin through two cycles of polymerization--depolymerization in a
  high-molarity buffer},\ }\href@noop {} {\bibfield  {journal} {\bibinfo
  {journal} {Protein Expression and Purification}\ }\textbf {\bibinfo {volume}
  {32}},\ \bibinfo {pages} {83 } (\bibinfo {year} {2003})}\BibitemShut
  {NoStop}%
\bibitem [{\citenamefont {Hyman}\ \emph {et~al.}(1991)\citenamefont {Hyman},
  \citenamefont {Drechsel}, \citenamefont {Kellogg}, \citenamefont {Salser},
  \citenamefont {Sawin}, \citenamefont {Steffen}, \citenamefont {Wordeman},\
  and\ \citenamefont {Mitchison}}]{HymanTubulin}%
  \BibitemOpen
  \bibfield  {author} {\bibinfo {author} {\bibfnamefont {A.}~\bibnamefont
  {Hyman}}, \bibinfo {author} {\bibfnamefont {D.}~\bibnamefont {Drechsel}},
  \bibinfo {author} {\bibfnamefont {D.}~\bibnamefont {Kellogg}}, \bibinfo
  {author} {\bibfnamefont {S.}~\bibnamefont {Salser}}, \bibinfo {author}
  {\bibfnamefont {K.}~\bibnamefont {Sawin}}, \bibinfo {author} {\bibfnamefont
  {P.}~\bibnamefont {Steffen}}, \bibinfo {author} {\bibfnamefont
  {L.}~\bibnamefont {Wordeman}},\ and\ \bibinfo {author} {\bibfnamefont
  {T.}~\bibnamefont {Mitchison}},\ }\bibfield  {title} {\bibinfo {title}
  {Preparation of modified tubulins},\ }in\ \href
  {https://doi.org/https://doi.org/10.1016/0076-6879(91)96041-O} {\emph
  {\bibinfo {booktitle} {Molecular Motors and the Cytoskeleton}}},\ \bibinfo
  {series} {Methods in Enzymology}, Vol.\ \bibinfo {volume} {196}\ (\bibinfo
  {publisher} {Academic Press},\ \bibinfo {year} {1991})\ pp.\ \bibinfo {pages}
  {478 -- 485}\BibitemShut {NoStop}%
\bibitem [{\citenamefont {Lau}\ \emph {et~al.}(2009)\citenamefont {Lau},
  \citenamefont {Prasad},\ and\ \citenamefont {Dogic}}]{Lau_2009}%
  \BibitemOpen
  \bibfield  {author} {\bibinfo {author} {\bibfnamefont {A.~W.~C.}\
  \bibnamefont {Lau}}, \bibinfo {author} {\bibfnamefont {A.}~\bibnamefont
  {Prasad}},\ and\ \bibinfo {author} {\bibfnamefont {Z.}~\bibnamefont
  {Dogic}},\ }\bibfield  {title} {\bibinfo {title} {Condensation of isolated
  semi-flexible filaments driven by depletion interactions},\ }\href@noop {}
  {\bibfield  {journal} {\bibinfo  {journal} {{EPL} (Europhysics Letters)}\
  }\textbf {\bibinfo {volume} {87}},\ \bibinfo {pages} {48006} (\bibinfo {year}
  {2009})}\BibitemShut {NoStop}%
\bibitem [{\citenamefont {Berg}\ \emph {et~al.}(2019)\citenamefont {Berg},
  \citenamefont {Kutra}, \citenamefont {Kroeger}, \citenamefont {Straehle},
  \citenamefont {Kausler}, \citenamefont {Haubold}, \citenamefont {Schiegg},
  \citenamefont {Ales}, \citenamefont {Beier}, \citenamefont {Rudy},
  \citenamefont {Eren}, \citenamefont {Cervantes}, \citenamefont {Xu},
  \citenamefont {Beuttenmueller}, \citenamefont {Wolny}, \citenamefont {Zhang},
  \citenamefont {Koethe}, \citenamefont {Hamprecht},\ and\ \citenamefont
  {Kreshuk}}]{ilastik}%
  \BibitemOpen
  \bibfield  {author} {\bibinfo {author} {\bibfnamefont {S.}~\bibnamefont
  {Berg}}, \bibinfo {author} {\bibfnamefont {D.}~\bibnamefont {Kutra}},
  \bibinfo {author} {\bibfnamefont {T.}~\bibnamefont {Kroeger}}, \bibinfo
  {author} {\bibfnamefont {C.~N.}\ \bibnamefont {Straehle}}, \bibinfo {author}
  {\bibfnamefont {B.~X.}\ \bibnamefont {Kausler}}, \bibinfo {author}
  {\bibfnamefont {C.}~\bibnamefont {Haubold}}, \bibinfo {author} {\bibfnamefont
  {M.}~\bibnamefont {Schiegg}}, \bibinfo {author} {\bibfnamefont
  {J.}~\bibnamefont {Ales}}, \bibinfo {author} {\bibfnamefont {T.}~\bibnamefont
  {Beier}}, \bibinfo {author} {\bibfnamefont {M.}~\bibnamefont {Rudy}},
  \bibinfo {author} {\bibfnamefont {K.}~\bibnamefont {Eren}}, \bibinfo {author}
  {\bibfnamefont {J.~I.}\ \bibnamefont {Cervantes}}, \bibinfo {author}
  {\bibfnamefont {B.}~\bibnamefont {Xu}}, \bibinfo {author} {\bibfnamefont
  {F.}~\bibnamefont {Beuttenmueller}}, \bibinfo {author} {\bibfnamefont
  {A.}~\bibnamefont {Wolny}}, \bibinfo {author} {\bibfnamefont
  {C.}~\bibnamefont {Zhang}}, \bibinfo {author} {\bibfnamefont
  {U.}~\bibnamefont {Koethe}}, \bibinfo {author} {\bibfnamefont {F.~A.}\
  \bibnamefont {Hamprecht}},\ and\ \bibinfo {author} {\bibfnamefont
  {A.}~\bibnamefont {Kreshuk}},\ }\bibfield  {title} {\bibinfo {title}
  {ilastik: interactive machine learning for (bio)image analysis},\ }\bibfield
  {journal} {\bibinfo  {journal} {Nature Methods}\ }\href
  {https://doi.org/10.1038/s41592-019-0582-9} 
  (\bibinfo {year} {2019})\BibitemShut {NoStop}%
\end{thebibliography}

\end{document}